\documentclass{article}

\usepackage{microtype}
\usepackage{graphicx}
\usepackage{subfigure}
\usepackage{booktabs} 

\usepackage{hyperref}

\usepackage[accepted]{icml2024}

\usepackage{amsmath}
\usepackage{amssymb}
\usepackage{mathtools}
\usepackage{amsthm}

\usepackage[capitalize,noabbrev]{cleveref}

\theoremstyle{plain}

\theoremstyle{definition}

\theoremstyle{remark}

\usepackage{stmaryrd}
\usepackage{algpseudocode}

\newcommand{\pixel}{\pmb{q}}
\newcommand{\Poipmf}[2]{\text{Poi}\left(#2\right)}
\newcommand{\kPoipmf}[3]{\text{TruncPoi}\left(#2, #3\right)}
\newcommand{\Gammapdf}[3]{\text{Gamma}\bigl(#2, #3\bigr)}
\newcommand{\nmax}{n_{\max}}
\renewcommand{\nmax}{{n^*}}

\newcommand{\poipmf}[2]{\text{Poi}\left(#1; #2\right)}

\newcommand{\kpoipmf}[3]{\text{TruncPoi}\left(#1; #2, #3\right)}
\newcommand{\gammapdf}[3]{\text{Gamma}\left(#1; #2, #3\right)}

\newcommand{\iverson}[1]{\llbracket #1 \rrbracket}

\icmltitlerunning{Bayesian MEF for robust HDR ptychography}

\begin{document}
\twocolumn[
\icmltitle{Bayesian multi-exposure image fusion for robust \\high dynamic range ptychography}
%
%
%
\icmlsetsymbol{equal}{*}
\begin{icmlauthorlist}
\icmlauthor{Shantanu Kodgirwar}{UKJ}
\icmlauthor{Lars Loetgering}{Zeiss}
\icmlauthor{Chang Liu}{HIJ,GSI,IAP}
\icmlauthor{Aleena Joseph}{IAP}
\icmlauthor{Leona Licht}{HIJ,GSI,IAP}
\icmlauthor{Daniel S. Penagos Molina}{HIJ,GSI,IAP}
\icmlauthor{Wilhelm Eschen}{HIJ,GSI,IAP}
\icmlauthor{Jan Rothhardt}{HIJ,GSI,IAP,IOF}
\icmlauthor{Michael Habeck}{UKJ,FMC,MPI}
\end{icmlauthorlist}
\vspace{1mm}
\begin{center}
	\begin{minipage}{0.83\textwidth}
		\raggedright
		\begin{tabular}{l}
			$^1$ Faculty of Medicine, University of Jena, 07743 Jena, Germany\\
			$^2$ ZEISS Research Microscopy Solutions, 07745 Jena, Germany\\
			$^3$ Helmholtz-Institute Jena, 07743 Jena, Germany\\
			$^4$ GSI Helmholtzzentrum für Schwerionenforschung, 64291 Darmstadt, Germany\\
			$^5$ Institute of Applied Physics and Abbe Center of Photonics, University of Jena,
			07745 Jena, Germany\\
			$^6$ Fraunhofer Institute for Applied Optics and Precision Engineering, 07745 Jena, Germany\\
			$^7$ Faculty of Mathematics and Computer Science, University of Jena, 07743 Jena, Germany\\
			$^8$ Max Planck Institute for Multidisciplinary Sciences, 37077 Goettingen, Germany\\ 
		\end{tabular}
	\end{minipage}
\end{center}

\icmlaffiliation{UKJ}{Faculty of Medicine, Friedrich Schiller University Jena, Germany}
\icmlaffiliation{Zeiss}{ZEISS Research Microscopy Solutions, Jena, Germany}
\icmlaffiliation{HIJ}{Helmholtz-Institute Jena, Jena, Germany}
\icmlaffiliation{GSI}{GSI Helmholtzzentrum für Schwerionenforschung, Darmstadt, Germany}
\icmlaffiliation{IAP}{Institute of Applied Physics and Abbe Center of Photonics, Friedrich Schiller University Jena, Jena, Germany}
\icmlaffiliation{IOF}{Fraunhofer IOF, Albert-Einstein-Strasse 7, 07745 Jena, Germany}
\icmlaffiliation{FMC}{Faculty of Mathematics and Computer Science, Friedrich Schiller University Jena, Jena, Germany}
\icmlaffiliation{MPI}{Max Plank Institute for Multidisciplinary Sciences, Am Fassberg 11, 37077 Goettingen, Germany}
\icmlcorrespondingauthor{Shantanu Kodgirwar}{shantanu.kodgirwar@uni-jena.de}
%
%
\vskip 0.3in
%
%
%
%
\printAffiliationsAndNotice{}  
%
\begin{abstract}
	The limited dynamic range of the detector can impede coherent diffractive imaging (CDI) schemes from achieving diffraction-limited resolution. To overcome this limitation, a straightforward approach is to utilize high dynamic range (HDR) imaging through multi-exposure image fusion (MEF). This method involves capturing measurements at different exposure times, spanning from under to overexposure and fusing them into a single HDR image. The conventional MEF technique in ptychography typically involves subtracting the background noise, ignoring the saturated pixels and then merging the acquisitions. However, this approach is inadequate under conditions of low signal-to-noise ratio (SNR). Additionally, variations in illumination intensity significantly affect the phase retrieval process. To address these issues, we propose a Bayesian MEF modeling approach based on a modified Poisson distribution that takes the background and saturation into account. To infer the model parameters, the expectation-maximization (EM) algorithm is employed. As demonstrated with synthetic and experimental data, our approach outperforms the conventional MEF method, offering superior phase retrieval under challenging experimental conditions. This work underscores the significance of robust multi-exposure image fusion for ptychography, particularly in imaging shot-noise-dominated weakly scattering specimens or in cases where access to HDR detectors with high SNR is limited. Furthermore, the applicability of the Bayesian MEF approach extends beyond CDI to any imaging scheme that requires HDR treatment. Given this versatility, we provide the implementation of our algorithm as a Python package.
\end{abstract}
]
\section{Introduction}
\label{sec:introduction}

\begin{figure}[htbp]
	\centering
  \includegraphics[width=\linewidth]{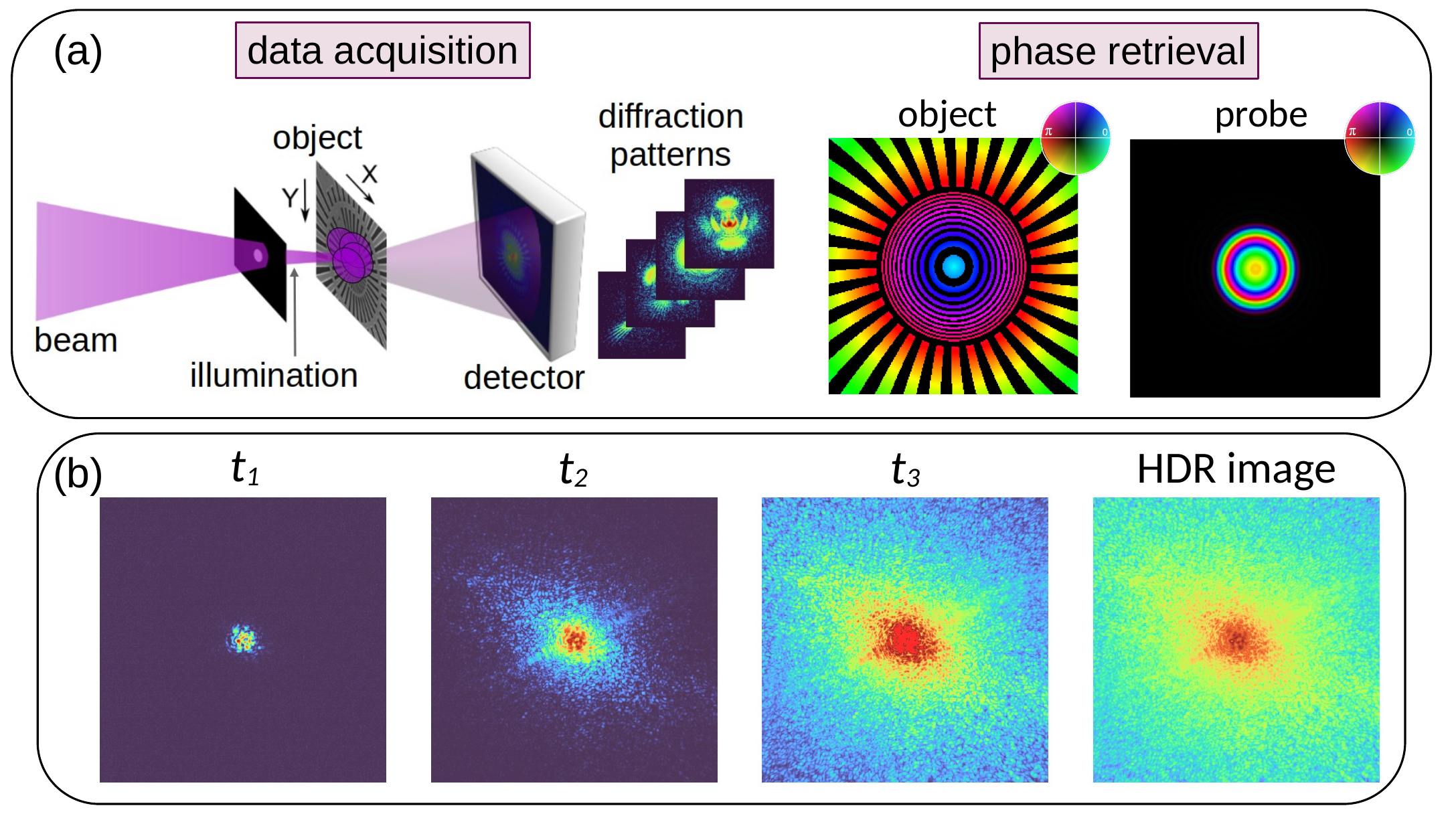}
  \caption{(a) Principle of conventional ptychography from data acquisition to phase retrieval. (b) Diffraction patterns are recorded at increasing exposure times (bright red pixels are saturated) and fused to achieve a high dynamic range (HDR) image.}
  \label{fig:schematic-mef}
\end{figure}

Ptychography~\cite{Maiden2009, Rodenburg2019}, a scanning coherent diffractive imaging (CDI) method, has gained popularity as a "lensless" computational imaging scheme allowing for simultaneous phase microscopy and wavefront sensing. This is especially advantageous in the short-wavelength regime~\cite{Thibault2008, Loetgering2022}, where the availability of high-quality imaging optics is scarce. This technique involves laterally moving a thin specimen across localized illumination while recording a series of diffraction patterns. By leveraging the overlap between adjacent scan positions, the captured diffraction data can be transformed into complex-valued reconstructions of the specimen and illumination wavefront using a phase-retrieval algorithm (Fig.~\ref{fig:schematic-mef}a). The camera exposure time is a critical parameter for the acquisition of diffraction patterns. Photons diffracted towards large angles tend to exhibit lower intensities, resulting in a poorer signal-to-noise ratio (SNR). These photons carry high-spatial-frequency information and are usually accessible through measurements with longer acquisition times. However, this may result in the saturation of the pixels at the center of the diffraction patterns due to the limited dynamic range of the detector. To overcome this limitation, high dynamic range (HDR) imaging can be employed as a preprocessing step in CDI before phase retrieval.

Multi-exposure image fusion (MEF) involves recording multiple acquisitions at varying exposure times and fusing them into a single HDR image. Many MEF algorithms have been proposed in the field of computer vision~\cite{Kede2015, Zhang2021, Fang2022}, with some adoption in microscopy~\cite{Yin2015, Singh2022, Liu2023}. However, to the best of our knowledge, these algorithms  have not been directly applied to CDI, primarily because they were not designed for reciprocal space data undergoing phase retrieval. Therefore, in CDI, a predominantly simplistic approach for HDR image fusion has been employed~\cite{Baksh2016, Rose2018, Lo2018}, referred to here as the {\em conventional MEF} method. In ptychography, multiple diffraction patterns per scan position are recorded with increasing exposure times and are merged together (see Fig.~\ref{fig:schematic-mef}b). The conventional MEF method is, however, inadequate in low signal-to-noise ratio (SNR) conditions. For example, diffraction patterns captured from weakly scattering specimens exhibit high levels of shot noise towards large scattering angles. This is due to the diffraction signal being dominated by ballistic (or unscattered) photons. Therefore, performing ptychographic phase retrieval for these specimens is challenging in general~\cite{Dierolf2010}. Usually, to overcome this and relax the dynamic range requirements, highly focused structured beams are utilized~\cite{Stockmar2013, Odstrcil2019, Eschen2024}. However, in extreme wavelengths such as the X-ray regime, acquiring high-NA focusing optics is difficult~\cite{Mimura2010, Bajt2018}, thereby lowering the efficacy of these wavefront modulation techniques. Moreover, specialized structuring elements must be added to the beam path~\cite{Eschen2022}. Therefore, MEF algorithms are instrumental in achieving HDR and maintaining high flexibility with regard to the design of the imaging experiments.
\par We briefly review the conventional MEF strategy and its limitations before presenting our solution. For every scan position, multiple diffraction patterns $n_i(\pixel)$ are recorded for varying acquisition times $t_i$ ($i=1, \ldots, K$), where $K$ denotes the number of acquisitions and $\pixel$ represents the pixel index. The background estimate $b_i(\pixel)$ characterizes camera readout noise. A simple correction for background noise is the subtraction from the diffraction patterns $n_i^+(\pixel) := \max{\left(0, n_i(\pixel) - b_i(\pixel)\right)}$, where the max operation enforces nonnegativity. For every acquisition time $t_i$, there is an associated flux factor $c_i \approx t_i$. These patterns are then fused together into an estimated diffraction intensity  $\mathcal{I}(\pixel)$ based on maximum likelihood estimation (MLE) under the assumption of Poisson noise (see supplementary section \ref{sec:B}):
\begin{equation}\label{eq:mef_mle}
  \begin{aligned}
    \mathcal{I}(\pixel) &= \frac{\sum_i w_i(\pixel) n_i^+(\pixel)}{\sum_i w_i(\pixel) c_i(\pixel)}, \\
    w_i(\pixel) &= 
    \begin{cases}
      1, & \text{$\pixel$ is not saturated} \\
      0, & \text{$\pixel$ is saturated}
    \end{cases}
  \end{aligned}
\end{equation}
Pixel-wise binary weight $w_i(\pixel)$ ensures that only unsaturated data are used for the fusion process. Even if the exposure time is accurately known, naive MLE estimation suffers from several shortcomings, especially under low-SNR conditions. The background subtraction procedure clips negative values to zero, resulting in a loss of information. The $\max$ operation also introduces nonlinearity into intensity estimation, leading to a potential bias~\cite{Seifert2023}. Additionally, this approach ignores saturated data, contributing to another source of information loss and leading to a suboptimal fusion result. Moreover, the flux factors $c_i$ may not be known accurately. In the case of probe intensity fluctuations~\cite{Odstrcil2016} or if the detector behaves nonlinearly under varying exposure times~\cite{Suhaidi2009}, the conventional approach relies on a heuristic estimate of the flux factors, which affects the quality of fusion. In addition, recognizing the advantage of flexibly modifying the optical intensity, especially by increasing it while adjusting the camera exposure times, can prove beneficial for faster data acquisition. In this case, the uncertainty in $c_i$ will increase, introducing systematic errors in the fused data, thereby significantly affecting phase retrieval. 
\par To tackle these issues and achieve robust HDR imaging in ptychography, we propose a principled Bayesian modeling approach to multi-exposure image fusion which we refer to here as the {\em Bayesian MEF} method. We start by explaining our probabilistic model, followed by the inference procedure via the expectation-maximization (EM)~\cite{Dempster1977} algorithm. We then demonstrate the efficacy of our approach on synthetic data under low SNR conditions, assuming fixed and accurately known flux factors. Subsequently, we show results from the experimental data in a high SNR, and high photon count setting, where the flux factors are only heuristically known.
\section{Bayesian MEF}
\label{sec:Bayesian-MEF}
At the core of a Bayesian approach is the likelihood, a probabilistic model for the data. Here, the model must account for two effects: (a) background noise and (b) saturation owing to overexposure. Suppose we ignore the background noise and saturation effects. In that case, an appropriate likelihood for the $i$-th exposure is a Poisson distribution ~\cite{Thibault2012} with mean $c_i \mathcal I(\pixel)$, i.e., 
\begin{equation}
	\label{eq:likelihood-standard}
	n_i(\pixel) \sim \Poipmf{n_i(\pixel)}{c_i \mathcal{I}(\pixel)}
\end{equation} 
where $\sim$ denotes "follows the distribution". The second component of the Bayesian model is the prior. We assume a Gamma distribution as a prior with shape and scale parameters $\alpha_{\mathcal I}$ and $\beta_{\mathcal I}$. This choice encodes the assumption that $\mathcal I$ is non-negative and constant on average, $\mathcal I(\pixel) \approx \alpha_{\mathcal I} / \beta_{\mathcal I}$, and the fluctuations in $\mathcal I$ are on the order of $\alpha_{\mathcal I}/\beta_{\mathcal I}^2$~\cite{Gelman2013}. The expectation for a uniform Gamma prior of $\mathcal{I}(\pixel)$ can be given as $\alpha_\mathcal{I}/\beta_\mathcal{I} = 1$. To achieve this, we simply provide the same value for $\alpha_\mathcal{I}$ and $\beta_\mathcal{I}$. Here, we choose both of these values to be $10^{-3}$, which also avoids instability in the division. In case we have more detailed prior knowledge about $\mathcal I$, it can be incorporated by letting $\alpha_{\mathcal I}$ and $\beta_{\mathcal I}$ depend on pixel location $\pixel$. We also assume gamma priors for the unknown flux factors $c_i$ where the hyperparameters $\alpha_i, \beta_i$ are typically chosen such that the expected flux factor equals the acquisition time $\alpha_i/\beta_i = t_i$. For the special case of $\alpha_i=1$ which is the exponential distribution, $\beta_i=t_i^{-1}$. The conditional posterior distributions for $\mathcal I(\pixel)$ and $c_i$ are also gamma distributions. The model parameters can be estimated by iterating over the updates
\begin{equation}\label{eq:mef_ICE}
        \begin{aligned}
		c_i^{(k+1)} &= \frac{\alpha_i + \sum_{\pixel} n_i(\pixel)}{\beta_i + \sum_{\pixel} \mathcal{I}^{(k)}(\pixel)}\,, \\
		\mathcal{I}^{(k+1)}(\pixel) 
		&= \frac{\alpha_\mathcal{I} + \sum_i n_i(\pixel)}{\beta_\mathcal{I} + \sum_i c_i^{(k+1)}}
        \end{aligned}
\end{equation}
with iteration index denoted by superscript $(k)$. The derivation of the updates in Eq.~(\ref{eq:mef_ICE}) are provided in supplementary section \ref{sec:C1}. For $\alpha_\mathcal{I}=\beta_\mathcal{I}=0$, the estimate for $\mathcal{I}(\pixel)$ matches the maximum-likelihood estimate for the unsaturated pixels in Eq.~(\ref{eq:mef_mle}). 

To model the readout noise, we consider the mean $c_i\mathcal{I}(\pixel)$ to be shifted by the background mean $b_i(\pixel)$ such that $n_i(\pixel) \sim \Poipmf{n_i(\pixel)}{c_i \mathcal{I}(\pixel) + b_i(\pixel)}$. Since the sum of two Poisson distributions is again a Poisson distribution with both means added together, the data can be interpreted as the sum of the background counts generated from $\Poipmf{}{b_i(\pixel)}$ and noisefree counts  $n_{\mathcal{I} i}(\pixel) \sim \Poipmf{}{c_i \mathcal{I}(\pixel)}$. Only the sum of the background and noisefree counts is observed; however, $n_{\mathcal Ii}(\pixel)$ can be estimated from $n_i(\pixel)$. The conditional posterior of the noisefree counts is a binomial distribution (see supplementary section \ref{sec:C2}), and their expected values are
\begin{equation}\label{eq:background-removal}
  \mathbb E[n_{\mathcal Ii}(\pixel)\mid \mathcal I(\pixel), c_i, b_i(\pixel)] = \frac{c_i \mathcal I(\pixel)}{c_i \mathcal I(\pixel) + b_i(\pixel)} n_i(\pixel)\, .
\end{equation}
These are fractions of the observed counts $n_i (\pixel)$, which differ from estimating $n_{\mathcal Ii} (\pixel)$ (or $n_i^+ (\pixel)$ defined earlier) that involves the harsh background subtraction procedure.

To take the saturated pixels into account, we assume that the data $n_i(\pixel)$ are {\em censored} at the detector threshold $\nmax$ meaning that they are clipped at a maximum value of $\nmax$. The likelihood is a {\em (right-) censored} Poisson distribution (i.e., the right tail of the Poisson distribution is not observed). Let $\nu_i(\pixel)$ denote the latent uncensored counts, $\nu_i(\pixel) \sim \Poipmf{\nu_i(\pixel)}{c_i\mathcal I(\pixel) + b_i(\pixel)}$, then the observed data can be modeled as $n_i(\pixel) = \nu_i(\pixel)$ if $\nu_i(\pixel) < \nmax$ (or equivalently, $w_i(\pixel) = 1$), and $n_i(\pixel) = \nmax$ if $\nu_i(\pixel) \ge \nmax$ (or $w_i(\pixel) = 0$). If we know $\nu_i(\pixel)$, then parameter estimation would simplify to background removal under the Poisson model, as in Eq. (\ref{eq:background-removal}). However, since we do not observe the uncensored counts directly, we have to estimate them. This can be achieved with the following conditional posterior distributions
\begin{equation}\label{eq:conditional-posterior}
  \begin{aligned}
    \nu_i(\pixel) 
    &\sim \begin{cases}
      \kPoipmf{\nu_i(\pixel)}{c_i \mathcal{I}(\pixel) + b_i(\pixel)}{\nmax-1}, & w_i(\pixel) = 0 \\
      \Poipmf{n_i(\pixel)}{c_i \mathcal{I}(\pixel) + b_i(\pixel)}, & w_i(\pixel) = 1
    \end{cases}\\
    \nu_{\mathcal{I} i}(\pixel) &\sim \text{Binomial}\left(\nu_i(\pixel), \frac{c_i\mathcal{I}(\pixel)}{c_i\mathcal{I}(\pixel) + b_i(\pixel)}\right) \\
    c_i &\sim \Gammapdf{c_i}{\alpha_i + \sum_{\pixel} \nu_{\mathcal{I} i}(\pixel)}{\beta_i + \sum_{\pixel} \mathcal{I} (\pixel)} \\
    \mathcal{I}(\pixel) &\sim  \Gammapdf{\mathcal{I}(\pixel)}{\alpha_\mathcal{I} + \sum_i \nu_{\mathcal{I} i}(\pixel)}{\beta_\mathcal{I} + \sum_i c_i}		
  \end{aligned}
\end{equation}
where $\nu_{\mathcal{I} i}(\pixel)$ are the noisefree contributions to the uncensored counts $\nu_i(\pixel)$ and $\text{TruncPoi}$ denotes the truncated Poisson distribution~\cite{Plackett1953}. The model parameters are inferred via the expectation-maximization (EM) algorithm, which is an iterative procedure for MLE or maximum a posteriori (MAP) estimation in the presence of incomplete data. The E-step gives the expectation of the latent counts $\nu_i$, allowing us to estimate the latent noisefree counts $\nu_{\mathcal{I} i}$. In the M-step, the estimate of our model parameters; the fused pattern $\mathcal{I}$ and flux factors $c_i$ are maximized (see supplementary section \ref{sec:C3} for details).

Our EM algorithm, referred to here as Bayesian MEF is summarized in algorithm~\ref{alg:em-ice}. The key steps are: Estimation of the uncensored counts $\nu_i$ via their expected values under the truncated Poisson distribution (line 4), background removal (line 5) analogous to Eq. (\ref{eq:background-removal}), updating the flux factors (line 6) and estimation of the fused pattern (line 7) analogous to Eq. (\ref{eq:mef_ICE}). This procedure iteratively estimates the saturated pixels, reduces the measured background noise, allows the correction of flux factors and fuses the acquisitions for HDR imaging. The implementation of this algorithm can be found as a python package in Ref.~\cite{BayesMEF_Software_2024}.
\begin{algorithm}[ht]
	\caption{Bayesian MEF}
	\label{alg:em-ice}
	\begin{algorithmic}[1]
	  \State Initialize $\alpha_\mathcal{I} = \beta_\mathcal{I} = 0.001$, $\alpha_i=1.0$ and $\beta_i=t_i^{-1}$
	  \State Initialize $c_i^{(0)} \!\!= t_i$, and $\mathcal{I}^{(0)} \!\!= \dfrac{\alpha_\mathcal{I} + \sum_i w_i n_i}{\beta_\mathcal{I} + \sum_i w_i t_i}$
	  \vspace{0.6mm}
	  \Repeat
	  \vspace{0.7mm}
	  \State $\nu_i^{(k+1)} \gets w_i n_i + \overline{w_i}\,\mathbb E[\nu_i | c_i^{(k)}\mathcal{I}^{(k)} + b_i, \nmax -1]$
	  \vspace{0.8mm}
	  \State $\nu_{\mathcal{I} i}^{(k+1)} \gets \dfrac{c_i^{(k)}\mathcal{I}^{(k)}}{c_i^{(k)}\mathcal{I}^{(k)} + \,\,b_i} \nu_i^{(k+1)}$
	  \vspace{0.8mm}
	  \State $c_i^{(k+1)} \gets \dfrac{\alpha_i + \sum_{\pixel} \nu_{\mathcal{I} i}^{(k+1)}(\pixel)}{\beta_i + \sum_{\pixel} \mathcal{I}^{(k)}(\pixel)}$
	  \vspace{0.8mm}
	  \State $\mathcal{I}^{(k+1)} \gets \dfrac{\alpha_\mathcal{I} + \sum_i \nu_{\mathcal{I} i}^{(k+1)}}{\beta_\mathcal{I} + \sum_i c_i^{(k+1)}}$
	  \vspace{0.7mm}
	  \State $k \gets k+1$
	  \Until{converge}
	\end{algorithmic}
  \end{algorithm}
\section{Low SNR synthetic data with known flux factors}
\label{sec:simulations}
\begin{figure*}[ht]
	\centering
	\includegraphics[width=0.95\linewidth]{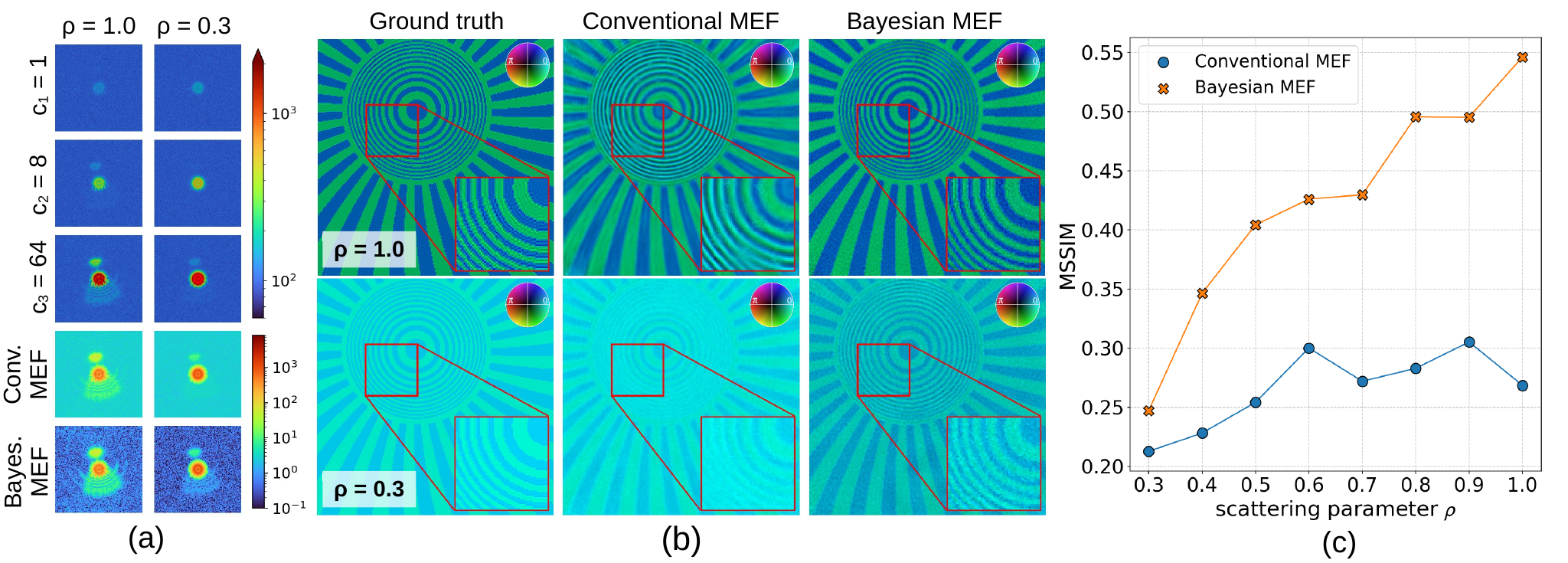}
	\caption{(a) Synthetic diffraction patterns (log-scale) for a particular scan position under strong scattering ($\rho=1.0$) and weak scattering ($\rho=0.3$) corresponding to different flux factors $c_i$. The third acquisition ($c_3 = 64.0$) exhibits saturation (bright red pixels). The data is then fused with conventional and Bayesian MEF methods. (b) Comparing ptychographic reconstructions (amplitude and phase) of the ground truth with MEF fused result from the synthetic data for $\rho=1.0$ (top-row) and $\rho=0.3$ (bottom-row). (c) Evaluating the mean structural similarity index (MSSIM) for reconstructed phase with data generated by varying $\rho$ and fused with the MEF methods.}
	\label{fig:synthetic-results}
\end{figure*}

To study the performance of Bayesian MEF, we considered synthetic data in a low-SNR setting. We fixed the flux factors $c_i = t_i$ and did not estimate them. To simulate low-SNR condition, we considered a weakly scattering specimen. Supplementary section \ref{sec:D1} derives the SNR condition under weak phase object approximation (WPOA)~\cite{TREACY2012}. In general, a phase object is given by $O(\pixel)= \exp{ \big(i \phi(\pixel)\big)}$ where $\rho \approx \max\big(\phi(\pixel)\big)$ denotes the maximum phase. Smaller values of $\rho$ result in weaker scattering, corresponding to a weakly phase-modulating specimen. To systematically study the impact on ptychographic reconstructions, we simulated the data by varying $\rho$ from weakly scattering $\rho=0.3$ to strongly scattering $\rho=1.0$. For a given $\rho$, the diffraction patterns are simulated using our forward model $n_i(\pixel) \sim \Poipmf{n_i(\pixel)}{c_i \mathcal{I}(\pixel) + b_i(\pixel)}$ where the additive background is drawn from the Gaussian distribution $b_i(\pixel) \sim \mathcal{N}(\mu_b, \sigma^2_b)$~\cite{Seifert2023}. Usually, the background counts are a small fraction of the actual data and remain fairly constant with low spatial variance. Therefore, we choose $\mu_b=100$ and $\sigma^2_b = 0.8$. We set $c_i$ to 1, 8, and 64, corresponding to maximum photon counts of $2^7, 2^{10}$ and $2^{13}$, respectively. The data was censored at a threshold of $2^{11}$ causing saturation mainly in the third acquisition. The simulated diffraction patterns are shown in Fig.~\ref{fig:synthetic-results}a, where acquisitions for a specific scan position were generated for $\rho=1.0$ and $\rho=0.3$. For the third acquisition ($c_3 = 64$), the number of saturated pixels was $\approx 2.3$ times higher for $\rho=0.3$ than for $\rho=1.0$, further increasing the level of uncertainty in the fusion of data from the weakly scattering specimen. For each exposure time, we repeated the measurement six times, fusing $18$ measurements per scan point. Utilizing fused data from all scan positions, we employed the momentum acceleration (mPIE)~\cite{Maiden2017} phase retrieval algorithm implemented in the open-source package \emph{PtyLab}~\cite{Loetgering2023}. Fig.~\ref{fig:synthetic-results}b compares the impact of the MEF methods on the reconstruction of a spokes target as our phase object. The reconstruction quality was significantly better with the Bayesian MEF method for strongly as well as weakly scattering data. Supplementary Fig. \ref{fig:synthetic-data} compares the results for $\rho$ varying from $0.3$ to $1.0$ showing improved reconstructions for all cases. These results are corroborated by evaluating the mean structural similarity (MSSIM) index~\cite{Wang2004} for phase reconstructions by varying $\rho$ (see Fig.~\ref{fig:synthetic-results}c). Supplementary section \ref{sec:D2} contains more details on the MSSIM metric as well as its use. As seen here, our method consistently shows higher MSSIM values for all values of $\rho$, whereas the improvement in the conventional case remains stagnant even for $\rho$ corresponding to strongly scattering ($\rho=1.0$) data.
\section{High SNR experimental data with heuristic flux factors}
\label{sec:experimental-demonstration}
The tests with synthetic data utilized the known flux factors $c_i$. However, these may or may not be known from an experiment. In the following experimental tests, we have a high SNR and high photon count setting; however, the flux factors are only heuristically known. We demonstrate that the fusion results from the conventional case become unreliable when the flux factors deviate significantly from the actual values. One such case is when the illumination intensity also varies with camera exposure time. In this case, we only have access to heuristic estimate $c_i^{heu}$. It can be calculated by summing the uncensored data for every acquisition and dividing it by the sum of uncensored data for a fixed acquisition. For example, out of $K$ acquisitions, the fixed acquisition could be the middle one $i=K/2$. The Bayesian MEF approach, outlined in algorithm~\ref{alg:em-ice}, should iteratively correct the heuristic flux factors by setting $c_i^{(0)} = c_i^{heu.}$. To test this, we constructed a conventional ptychography setup in transmission geometry. A coherent laser source from SuperK COMPACT, offering manual adjustment of the laser power in percentage was employed. We chose a wavelength of $625 \,nm$, a probe size of $400\,\mu m$, and used an unstained histological mouse section as our test object. Diffraction data was recorded at $500$ scan positions using a Lucid CMOS camera with a $12$ bit analog-to-digital converter (ADC) and a binning factor of $8$, allowing a maximum of $2^{18}$ counts. For additional experimental details, please refer to supplementary section \ref{sec:E}.

\begin{figure}[h]
	\centering
	\includegraphics[width=\linewidth]{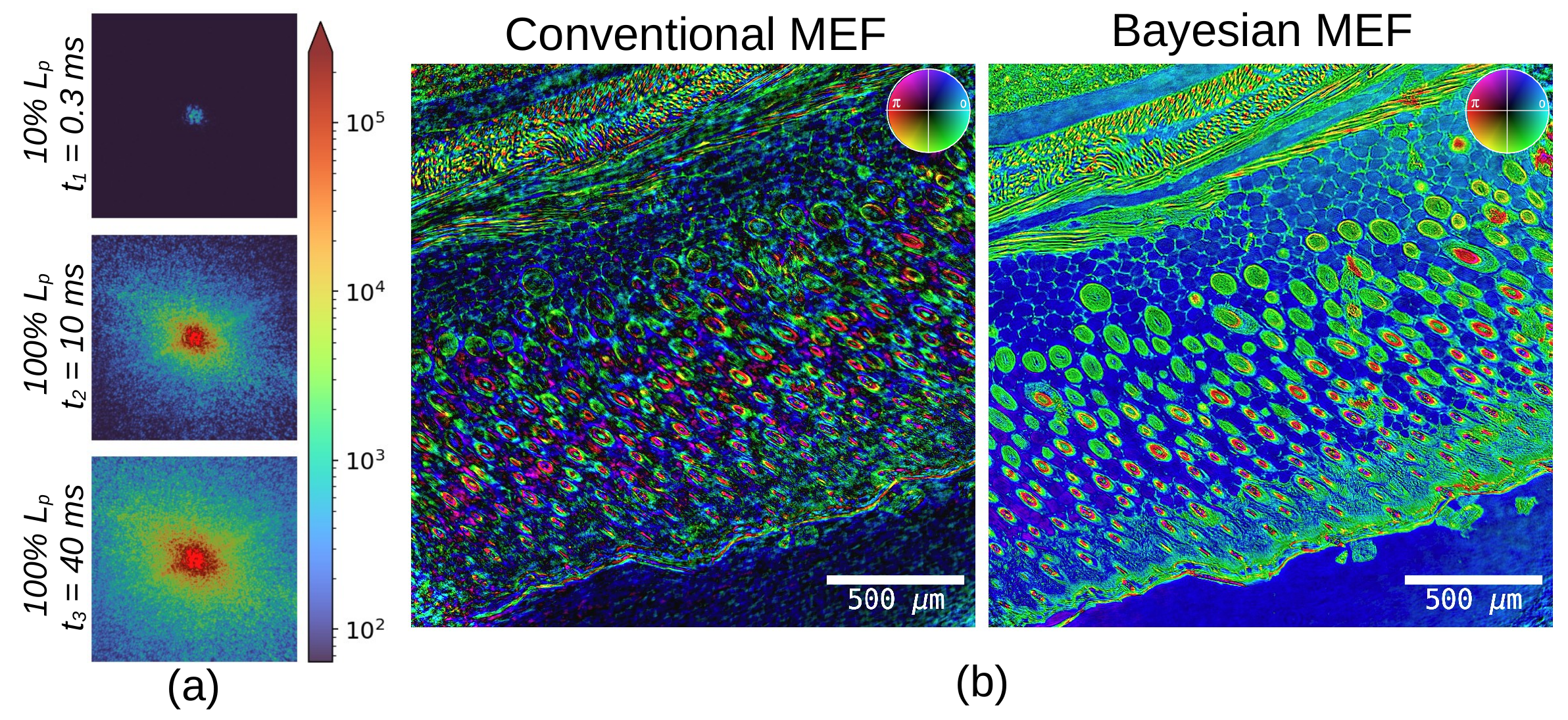}
	\caption{\textbf{(a)} Diffraction patterns recorded with varying laser power $L_p$ in percentage and camera acquisition times $t_i$. The second and the third acquisitions are overexposed. \textbf{(b)} Ptychographic reconstructions using the two MEF methods.}
	\label{fig:experimental-results}
\end{figure}

\begin{figure}[h]
	\centering
	\includegraphics[width=\linewidth]{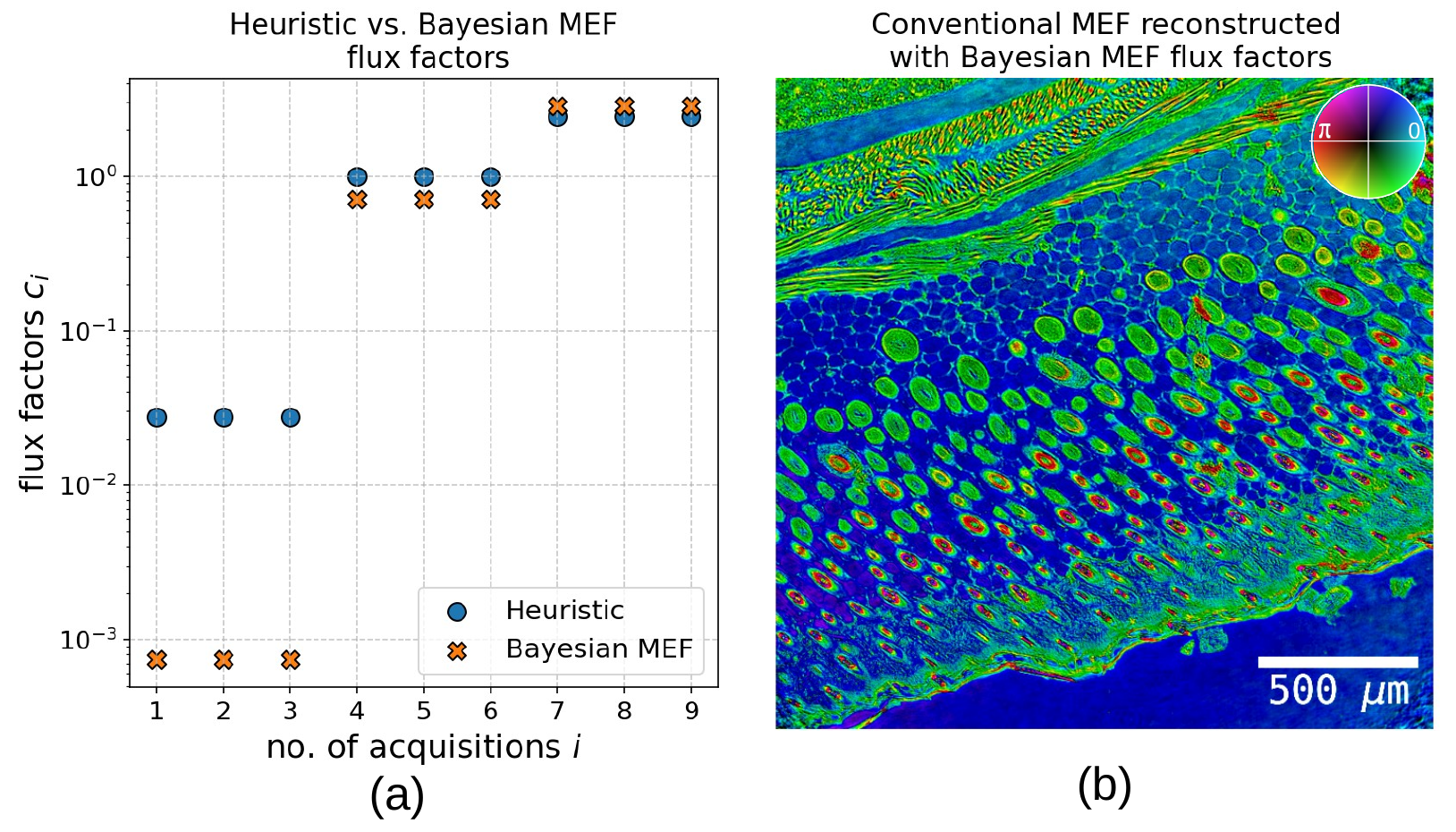}
	\caption{(a) Heuristic and the corrected Bayesian MEF flux factors (shown in log-scale) for the nine measurements (each acquisition repeated thrice) for every scan position. (b) Conventional MEF reconstruction that used the corrected Bayesian MEF flux factors.}
	\label{fig:heu_em_fluxes}
\end{figure}

We recorded three acquisitions by varying the exposure time and the laser power in percentage (refer to Fig.~\ref{fig:experimental-results}a). Each acquisition was repeated thrice, necessitating the fusion of nine diffraction patterns per scan. For each acquisition, $150$ dark-frame measurements were recorded and averaged. For the conventional MEF case, we applied the background subtraction procedure to the diffraction patterns and estimated the flux factors heuristically. Subsequently, the conventional MEF method was applied, resulting in a largely unsuccessful ptychographic reconstruction, as depicted in the left panel of Fig.~\ref{fig:experimental-results}b. However, our method corrects the heuristic flux factors and estimates fused patterns, leading to a high-quality ptychographic reconstruction (see Fig.~\ref{fig:experimental-results}b, right panel). It is expected that both MEF methods yield comparable results in a high SNR and high photon count scenario, where flux factors are accurately known. This is verified by utilizing the corrected fluxes from our method and using them for the conventional MEF case, leading to visually indistinguishable reconstructions (see Fig. \ref{fig:heu_em_fluxes}).
\section{Summary}
\label{sec:summary}
We introduced a Bayesian approach for MEF that incorporates image formation statistics with priors in a principled manner for its application in CDI. Our findings illustrate enhanced HDR phase retrieval, particularly in scenarios, characterized by low SNR or variations in illumination intensity. It can also be noted that conventional MEF is merely a special case of the Bayesian MEF approach, which was verified with the experimental data where both methods perform equivalently with known flux factors in a high count, high SNR regime. Currently, MEF is a preprocessing step for ptychography, and a natural extension of our work would involve the integration of our Bayesian model directly into a ptychography framework based on automatic differentiation (AD)~\cite{Kandel2019}, potentially further enhancing robustness in HDR ptychographic reconstructions. Furthermore, exploring physics-informed priors could enhance estimation in scenarios with limited data in the low SNR regime or expedite algorithmic convergence.

In summary, the results presented in this study signify noteworthy advancements in the field of computational imaging. Our work holds promise for robust phase retrieval with challenging weakly scattering specimens or when access to high-quality detectors at extreme wavelengths~\cite{Thibault2008, Loetgering2022} is restricted. Moreover, the versatility of our approach renders it applicable to any imaging scheme requiring enhancement of the dynamic range.

\section*{Acknowledgements}
\label{sec:funding}
SK and MH acknowledge funding by the Carl Zeiss Foundation within the program ``CZS Stiftungsprofessuren'' and by the German Research Foundation (DFG) within grant HA 5918/4-1. The remaining authors acknowledge support from the Helmholtz Association (ZT-I-PF-4-018 AsoftXm), the Free State of Thuringia and the European Social Fund Plus (2023 FGR0053). We acknowledge support by the German Research Foundation Projekt-Nr. 512648189 and the Open Access Publication Fund of the Thueringer Universitaets- und Landesbibliothek Jena.

\section*{Data Availability Statement}
In addition to the software package available in Ref.~\cite{BayesMEF_Software_2024}, the data used in this work is also available in Ref.~\cite{BayesMEF_data_2024}. To use the data and reproduce the results in this work, please refer to the details in the GitHub repository of the software.

\bibliography{references}

\begin{thebibliography}{35}
\providecommand{\natexlab}[1]{#1}
\providecommand{\url}[1]{\texttt{#1}}
\expandafter\ifx\csname urlstyle\endcsname\relax
  \providecommand{\doi}[1]{doi: #1}\else
  \providecommand{\doi}{doi: \begingroup \urlstyle{rm}\Url}\fi

\bibitem[Bajt et~al.(2018)Bajt, Prasciolu, Fleckenstein, Domarack{\'y}, Chapman, Morgan, Yefanov, Messerschmidt, Du, Murray, Mariani, Kuhn, Aplin, Pande, Villanueva-Perez, Stachnik, Chen, Andrejczuk, Meents, Burkhardt, Pennicard, Huang, Yan, Nazaretski, Chu, and Hamm]{Bajt2018}
Bajt, S., Prasciolu, M., Fleckenstein, H., Domarack{\'y}, M., Chapman, H.~N., Morgan, A.~J., Yefanov, O., Messerschmidt, M., Du, Y., Murray, K.~T., Mariani, V., Kuhn, M., Aplin, S., Pande, K., Villanueva-Perez, P., Stachnik, K., Chen, J.~P., Andrejczuk, A., Meents, A., Burkhardt, A., Pennicard, D., Huang, X., Yan, H., Nazaretski, E., Chu, Y.~S., and Hamm, C.~E.
\newblock X-ray focusing with efficient high-na multilayer laue lenses.
\newblock \emph{Light: Science {\&} Applications}, 7\penalty0 (3):\penalty0 17162--17162, Mar 2018.
\newblock ISSN 2047-7538.
\newblock \doi{10.1038/lsa.2017.162}.

\bibitem[Baksh et~al.(2016)Baksh, Odstr\v{c}il, Kim, Boden, Frey, and Brocklesby]{Baksh2016}
Baksh, P.~D., Odstr\v{c}il, M., Kim, H.-S., Boden, S.~A., Frey, J.~G., and Brocklesby, W.~S.
\newblock Wide-field broadband extreme ultraviolet transmission ptychography using a high-harmonic source.
\newblock \emph{Opt. Lett.}, 41\penalty0 (7):\penalty0 1317--1320, Apr 2016.
\newblock \doi{10.1364/OL.41.001317}.

\bibitem[Dempster et~al.(1977)Dempster, Laird, and Rubin]{Dempster1977}
Dempster, A.~P., Laird, N.~M., and Rubin, D.~B.
\newblock Maximum likelihood from incomplete data via the em algorithm.
\newblock \emph{Journal of the Royal Statistical Society: Series B (Methodological)}, 39\penalty0 (1):\penalty0 1--22, 1977.
\newblock \doi{https://doi.org/10.1111/j.2517-6161.1977.tb01600.x}.

\bibitem[Dierolf et~al.(2010)Dierolf, Thibault, Menzel, Kewish, Jefimovs, Schlichting, von König, Bunk, and Pfeiffer]{Dierolf2010}
Dierolf, M., Thibault, P., Menzel, A., Kewish, C.~M., Jefimovs, K., Schlichting, I., von König, K., Bunk, O., and Pfeiffer, F.
\newblock Ptychographic coherent diffractive imaging of weakly scattering specimens.
\newblock \emph{New Journal of Physics}, 12\penalty0 (3):\penalty0 035017, mar 2010.
\newblock \doi{10.1088/1367-2630/12/3/035017}.

\bibitem[Eschen et~al.(2022)Eschen, Loetgering, Schuster, Klas, Kirsche, Berthold, Steinert, Pertsch, Gross, Krause, Limpert, and Rothhardt]{Eschen2022}
Eschen, W., Loetgering, L., Schuster, V., Klas, R., Kirsche, A., Berthold, L., Steinert, M., Pertsch, T., Gross, H., Krause, M., Limpert, J., and Rothhardt, J.
\newblock Material-specific high-resolution table-top extreme ultraviolet microscopy.
\newblock \emph{Light: Science {\&} Applications}, 11\penalty0 (1):\penalty0 117, Apr 2022.
\newblock ISSN 2047-7538.
\newblock \doi{10.1038/s41377-022-00797-6}.

\bibitem[Eschen et~al.(2024)Eschen, Liu, Steinert, Molina, Siefke, Zeitner, Kaspar, Pertsch, Limpert, and Rothhardt]{Eschen2024}
Eschen, W., Liu, C., Steinert, M., Molina, D. S.~P., Siefke, T., Zeitner, U.~D., Kaspar, J., Pertsch, T., Limpert, J., and Rothhardt, J.
\newblock Structured illumination ptychography and at-wavelength characterization with an euv diffuser at 13.5 nm wavelength.
\newblock \emph{Opt. Express}, 32\penalty0 (3):\penalty0 3480--3491, Jan 2024.
\newblock \doi{10.1364/OE.507715}.

\bibitem[Gelman et~al.(2013)Gelman, Carlin, Stern, Dunson, Vehtari, and Rubin]{Gelman2013}
Gelman, A., Carlin, J.~B., Stern, H.~S., Dunson, D.~B., Vehtari, A., and Rubin, D.~B.
\newblock \emph{Bayesian Data Analysis}.
\newblock Chapman and Hall/CRC, November 2013.

\bibitem[Kandel et~al.(2019)Kandel, Maddali, Allain, Hruszkewycz, Jacobsen, and Nashed]{Kandel2019}
Kandel, S., Maddali, S., Allain, M., Hruszkewycz, S.~O., Jacobsen, C., and Nashed, Y. S.~G.
\newblock Using automatic differentiation as a general framework for ptychographic reconstruction.
\newblock \emph{Opt. Express}, 27\penalty0 (13):\penalty0 18653--18672, Jun 2019.
\newblock \doi{10.1364/OE.27.018653}.

\bibitem[Kodgirwar et~al.(2024{\natexlab{a}})Kodgirwar, Loetgering, Chang, Joseph, Molina, Eschen, Rothhardt, and Habeck]{BayesMEF_Software_2024}
Kodgirwar, S., Loetgering, L., Chang, L., Joseph, A., Molina, D. S.~P., Eschen, W., Rothhardt, J., and Habeck, M.
\newblock {Supplementary software: Bayesian multi-exposure image fusion for robust high dynamic range ptychography}, May 2024{\natexlab{a}}.
\newblock \href{https://doi.org/10.5281/zenodo.11103004}{10.5281/zenodo.11103004}.

\bibitem[Kodgirwar et~al.(2024{\natexlab{b}})Kodgirwar, Loetgering, Chang, Joseph, Molina, Wilhelm, Rothhardt, and Habeck]{BayesMEF_data_2024}
Kodgirwar, S., Loetgering, L., Chang, L., Joseph, A., Molina, D. S.~P., Wilhelm, E., Rothhardt, J., and Habeck, M.
\newblock {Supplementary data: Bayesian multi-exposure image fusion for robust high dynamic range ptychography}, April 2024{\natexlab{b}}.
\newblock \href{https://doi.org/10.5281/zenodo.10964223}{10.5281/zenodo.10964223}.

\bibitem[Liu et~al.(2023)Liu, Liu, Zou, Zhao, and Liu]{Liu2023}
Liu, J., Liu, C., Zou, C., Zhao, Y., and Liu, J.
\newblock Large dynamic range dark-field imaging based on microscopic images fusion.
\newblock \emph{Optics Communications}, 528:\penalty0 128966, 2023.
\newblock ISSN 0030-4018.
\newblock \doi{https://doi.org/10.1016/j.optcom.2022.128966}.

\bibitem[Lo et~al.(2018)Lo, Zhao, Gallagher-Jones, Rana, J.~Lodico, Xiao, Regan, and Miao]{Lo2018}
Lo, Y.~H., Zhao, L., Gallagher-Jones, M., Rana, A., J.~Lodico, J., Xiao, W., Regan, B.~C., and Miao, J.
\newblock In situ coherent diffractive imaging.
\newblock \emph{Nature Communications}, 9\penalty0 (1):\penalty0 1826, May 2018.
\newblock ISSN 2041-1723.
\newblock \doi{10.1038/s41467-018-04259-9}.

\bibitem[Loetgering et~al.(2022)Loetgering, Witte, and Rothhardt]{Loetgering2022}
Loetgering, L., Witte, S., and Rothhardt, J.
\newblock Advances in laboratory-scale ptychography using high harmonic sources [invited].
\newblock \emph{Opt. Express}, 30\penalty0 (3):\penalty0 4133--4164, Jan 2022.
\newblock \doi{10.1364/OE.443622}.

\bibitem[Loetgering et~al.(2023)Loetgering, Du, Flaes, Aidukas, Wechsler, Molina, Rose, Pelekanidis, Eschen, Hess, Wilhein, Heintzmann, Rothhardt, and Witte]{Loetgering2023}
Loetgering, L., Du, M., Flaes, D.~B., Aidukas, T., Wechsler, F., Molina, D. S.~P., Rose, M., Pelekanidis, A., Eschen, W., Hess, J., Wilhein, T., Heintzmann, R., Rothhardt, J., and Witte, S.
\newblock Ptylab.m/py/jl: a cross-platform, open-source inverse modeling toolbox for conventional and fourier ptychography.
\newblock \emph{Opt. Express}, 31\penalty0 (9):\penalty0 13763--13797, Apr 2023.
\newblock \doi{10.1364/OE.485370}.

\bibitem[Ma \& Wang(2015)Ma and Wang]{Kede2015}
Ma, K. and Wang, Z.
\newblock Multi-exposure image fusion: A patch-wise approach.
\newblock In \emph{2015 IEEE International Conference on Image Processing (ICIP)}, pp.\  1717--1721, 2015.
\newblock \doi{10.1109/ICIP.2015.7351094}.

\bibitem[Maiden et~al.(2017)Maiden, Johnson, and Li]{Maiden2017}
Maiden, A., Johnson, D., and Li, P.
\newblock Further improvements to the ptychographical iterative engine.
\newblock \emph{Optica}, 4\penalty0 (7):\penalty0 736--745, Jul 2017.
\newblock \doi{10.1364/OPTICA.4.000736}.

\bibitem[Maiden \& Rodenburg(2009)Maiden and Rodenburg]{Maiden2009}
Maiden, A.~M. and Rodenburg, J.~M.
\newblock An improved ptychographical phase retrieval algorithm for diffractive imaging.
\newblock \emph{Ultramicroscopy}, 109\penalty0 (10):\penalty0 1256--1262, 2009.
\newblock ISSN 0304-3991.
\newblock \doi{https://doi.org/10.1016/j.ultramic.2009.05.012}.

\bibitem[Mimura et~al.(2010)Mimura, Handa, Kimura, Yumoto, Yamakawa, Yokoyama, Matsuyama, Inagaki, Yamamura, Sano, Tamasaku, Nishino, Yabashi, Ishikawa, and Yamauchi]{Mimura2010}
Mimura, H., Handa, S., Kimura, T., Yumoto, H., Yamakawa, D., Yokoyama, H., Matsuyama, S., Inagaki, K., Yamamura, K., Sano, Y., Tamasaku, K., Nishino, Y., Yabashi, M., Ishikawa, T., and Yamauchi, K.
\newblock Breaking the 10{\thinspace}nm barrier in hard-x-ray focusing.
\newblock \emph{Nature Physics}, 6\penalty0 (2):\penalty0 122--125, Feb 2010.
\newblock ISSN 1745-2481.
\newblock \doi{10.1038/nphys1457}.

\bibitem[Odstrcil et~al.(2016)Odstrcil, Baksh, Boden, Card, Chad, Frey, and Brocklesby]{Odstrcil2016}
Odstrcil, M., Baksh, P., Boden, S.~A., Card, R., Chad, J.~E., Frey, J.~G., and Brocklesby, W.~S.
\newblock Ptychographic coherent diffractive imaging with orthogonal probe relaxation.
\newblock \emph{Opt. Express}, 24\penalty0 (8):\penalty0 8360--8369, Apr 2016.
\newblock \doi{10.1364/OE.24.008360}.

\bibitem[Odstr\v{c}il et~al.(2019)Odstr\v{c}il, Lebugle, Guizar-Sicairos, David, and Holler]{Odstrcil2019}
Odstr\v{c}il, M., Lebugle, M., Guizar-Sicairos, M., David, C., and Holler, M.
\newblock Towards optimized illumination for high-resolution ptychography.
\newblock \emph{Opt. Express}, 27\penalty0 (10):\penalty0 14981--14997, May 2019.
\newblock \doi{10.1364/OE.27.014981}.

\bibitem[Plackett(1953)]{Plackett1953}
Plackett, R.~L.
\newblock The truncated poisson distribution.
\newblock \emph{Biometrics}, 9\penalty0 (4):\penalty0 485, December 1953.

\bibitem[Rodenburg \& Maiden(2019)Rodenburg and Maiden]{Rodenburg2019}
Rodenburg, J. and Maiden, A.
\newblock \emph{Ptychography}, pp.\  819--904.
\newblock Springer International Publishing, 2019.
\newblock ISBN 978-3-030-00069-1.
\newblock \doi{10.1007/978-3-030-00069-1_17}.

\bibitem[Rose et~al.(2018)Rose, Senkbeil, von Gundlach, Stuhr, Rumancev, Dzhigaev, Besedin, Skopintsev, Loetgering, Viefhaus, Rosenhahn, and Vartanyants]{Rose2018}
Rose, M., Senkbeil, T., von Gundlach, A.~R., Stuhr, S., Rumancev, C., Dzhigaev, D., Besedin, I., Skopintsev, P., Loetgering, L., Viefhaus, J., Rosenhahn, A., and Vartanyants, I.~A.
\newblock Quantitative ptychographic bio-imaging in the water window.
\newblock \emph{Opt. Express}, 26\penalty0 (2):\penalty0 1237--1254, Jan 2018.
\newblock \doi{10.1364/OE.26.001237}.

\bibitem[Seifert et~al.(2023)Seifert, Shao, van Dam, Bouchet, van Leeuwen, and Mosk]{Seifert2023}
Seifert, J., Shao, Y., van Dam, R., Bouchet, D., van Leeuwen, T., and Mosk, A.~P.
\newblock Maximum-likelihood estimation in ptychography in the presence of poisson--gaussian noise statistics.
\newblock \emph{Opt. Lett.}, 48\penalty0 (22):\penalty0 6027--6030, Nov 2023.
\newblock \doi{10.1364/OL.502344}.

\bibitem[Shafie et~al.(2009)Shafie, Kawahito, Halin, and Hasan]{Suhaidi2009}
Shafie, S., Kawahito, S., Halin, I.~A., and Hasan, W. Z.~W.
\newblock Non-linearity in wide dynamic range {CMOS} image sensors utilizing a partial charge transfer technique.
\newblock \emph{Sensors (Basel)}, 9\penalty0 (12):\penalty0 9452--9467, nov 2009.

\bibitem[Singh et~al.(2022)Singh, Cristobal, Bueno, Blanco, Singh, Hrisheekesha, and Mittal]{Singh2022}
Singh, H., Cristobal, G., Bueno, G., Blanco, S., Singh, S., Hrisheekesha, P., and Mittal, N.
\newblock Multi-exposure microscopic image fusion-based detail enhancement algorithm.
\newblock \emph{Ultramicroscopy}, 236:\penalty0 113499, 2022.
\newblock ISSN 0304-3991.
\newblock \doi{https://doi.org/10.1016/j.ultramic.2022.113499}.

\bibitem[Stockmar et~al.(2013)Stockmar, Cloetens, Zanette, Enders, Dierolf, Pfeiffer, and Thibault]{Stockmar2013}
Stockmar, M., Cloetens, P., Zanette, I., Enders, B., Dierolf, M., Pfeiffer, F., and Thibault, P.
\newblock Near-field ptychography: phase retrieval for inline holography using a structured illumination.
\newblock \emph{Scientific Reports}, 3, May 2013.

\bibitem[Thibault \& Guizar-Sicairos(2012)Thibault and Guizar-Sicairos]{Thibault2012}
Thibault, P. and Guizar-Sicairos, M.
\newblock Maximum-likelihood refinement for coherent diffractive imaging.
\newblock \emph{New Journal of Physics}, 14\penalty0 (6):\penalty0 063004, jun 2012.
\newblock \doi{10.1088/1367-2630/14/6/063004}.

\bibitem[Thibault et~al.(2008)Thibault, Dierolf, Menzel, Bunk, David, and Pfeiffer]{Thibault2008}
Thibault, P., Dierolf, M., Menzel, A., Bunk, O., David, C., and Pfeiffer, F.
\newblock High-resolution scanning x-ray diffraction microscopy.
\newblock \emph{Science}, 321\penalty0 (5887):\penalty0 379--382, 2008.
\newblock \doi{10.1126/science.1158573}.

\bibitem[Treacy \& {Van Dyck}(2012)Treacy and {Van Dyck}]{TREACY2012}
Treacy, M. and {Van Dyck}, D.
\newblock A surprise in the first born approximation for electron scattering.
\newblock \emph{Ultramicroscopy}, 119:\penalty0 57--62, 2012.
\newblock ISSN 0304-3991.
\newblock \doi{https://doi.org/10.1016/j.ultramic.2011.11.012}.
\newblock Special Issue: Gertrude F. Rempfer 100th Birthday Memorial.

\bibitem[Van~der Walt et~al.(2014)Van~der Walt, Sch{\"o}nberger, Nunez-Iglesias, Boulogne, Warner, Yager, Gouillart, and Yu]{van2014scikit}
Van~der Walt, S., Sch{\"o}nberger, J.~L., Nunez-Iglesias, J., Boulogne, F., Warner, J.~D., Yager, N., Gouillart, E., and Yu, T.
\newblock scikit-image: image processing in python.
\newblock \emph{PeerJ}, 2:\penalty0 e453, 2014.

\bibitem[Wang et~al.(2004)Wang, Bovik, Sheikh, and Simoncelli]{Wang2004}
Wang, Z., Bovik, A., Sheikh, H., and Simoncelli, E.
\newblock Image quality assessment: from error visibility to structural similarity.
\newblock \emph{IEEE Transactions on Image Processing}, 13\penalty0 (4):\penalty0 600--612, 2004.
\newblock \doi{10.1109/TIP.2003.819861}.

\bibitem[Xu et~al.(2022)Xu, Liu, Song, Sun, and Wang]{Fang2022}
Xu, F., Liu, J., Song, Y., Sun, H., and Wang, X.
\newblock Multi-exposure image fusion techniques: A comprehensive review.
\newblock \emph{Remote Sensing}, 14\penalty0 (3), 2022.
\newblock ISSN 2072-4292.
\newblock \doi{10.3390/rs14030771}.

\bibitem[Yin et~al.(2015)Yin, Su, Ker, Li, and Li]{Yin2015}
Yin, Z., Su, H., Ker, E., Li, M., and Li, H.
\newblock Cell-sensitive phase contrast microscopy imaging by multiple exposures.
\newblock \emph{Medical Image Analysis}, 25\penalty0 (1):\penalty0 111--121, 2015.
\newblock ISSN 1361-8415.
\newblock \doi{https://doi.org/10.1016/j.media.2015.04.011}.

\bibitem[Zhang(2021)]{Zhang2021}
Zhang, X.
\newblock Benchmarking and comparing multi-exposure image fusion algorithms.
\newblock \emph{Information Fusion}, 74:\penalty0 111--131, 2021.
\newblock ISSN 1566-2535.
\newblock \doi{https://doi.org/10.1016/j.inffus.2021.02.005}.

\end{thebibliography}
\bibliographystyle{icml2024}


\appendix
\onecolumn
\section{Mathematical Notation}
\label{sec:A}

\begin{itemize}
	\item Iverson bracket:
	\[
	\iverson{A} = \begin{cases}
		1, & \text{$A$ is true} \\
		0, & \text{$A$ is false}
	\end{cases}
	\]
	\item Expectation:
	\[
	\mathbb E[f] = \int f(x)\, p(x)\, dx
	\]
	is the expected value of quantity $f$ under the probability distribution $p$.
	Special cases are the mean
	\[
	\mathbb E[x] = \int x\, p(x)\, dx
	\]
	and variance
	\[
	\text{var}[x] = \mathbb E\big[(x - \mathbb E[x])^2\big] = \mathbb E[x^2] - \left(\mathbb E[x]\right)^2\, .
	\]
	If $p(x \mid \theta)$ is a conditional probability distribution with parameter(s) $\theta$, then the expectations depend on $\theta$ which we make explicit by the following notation:
	\[
	\mathbb E[f\mid \theta] = \int f(x)\, p(x\mid\theta)\, dx
	\]
	\item ``Follows'':
	The tilde symbol ``$\sim$'' indicates that a quantity $x$ follows a particular probability distribution $p$:
	\[
	x \sim p(x)
	\]
	\item Parameters of the image model:
	\begin{description} 
		\item[$\pixel$:] detector pixel coordinates
		\item[$t_i$:] camera acquisition times (with $i=1, \ldots, K$ where $K$ is the number of acquisitions)
		\item[$n_i(\pixel)$:] diffraction patterns acquired during $t_i$ for a given scan position
		\item[$b_i(\pixel)$:] dark frame measurements acquired during $t_i$
		\item[$\mathcal{I}(\pixel)$:] fused image / diffraction pattern
		\item[$w_i(\pixel)$:] binary weight that is equal to $1$ for unsaturated pixels, $0$ otherwise
		\item[$c_i$:] flux rate associated with acquisition time $t_i$
	\end{description}
\end{itemize}

\begin{itemize}
	\item Poisson distribution:
	\[
	\poipmf{n}{\lambda} = \frac{\lambda^n}{n!} e^{-\lambda}, \quad \lambda > 0, \quad n=0, 1, 2, \ldots
	\]
	with {\em rate} $\lambda$ for counts $n$.
	\item $k$-truncated Poisson distribution:
	\[
	\kpoipmf{n}{\lambda}{k} \propto \poipmf{n}{\lambda}\, \iverson{n > k}
	\]
	This is a Poisson distribution with rate $\lambda$ where the counts are $n=k+1, k+2, \ldots$.
	The normalization constant is:
	\[
	\sum_{n>k} \poipmf{n}{\lambda} = 1 - \sum_{n\le k} \poipmf{n}{\lambda} = 1 - e^{-\lambda} \sum_{n=0}^k \frac{\lambda^n}{n!}\, .
	\]
	\item Gamma distribution:
	\[
	\gammapdf{x}{\alpha}{\beta} = \frac{\beta^\alpha}{\Gamma(\alpha)} x^{\alpha - 1} e^{-\beta x}, \quad x, \alpha, \beta \ge 0
	\]
	where $\Gamma(\cdot)$ is the Gamma function.
	The parameter $\alpha$ is the {\em shape parameter}, $\beta$ is the {\em rate parameter}.
	The mean and variance of a Gamma variable are:
	\[
	\mathbb E[x] = \frac{\alpha}{\beta}, \quad \text{var}[x] = \frac{\alpha}{\beta^2}\, .
	\]
	\item Binomial distribution:
	\[
	\text{Binomial}(k; n, p)
	= \binom{n}{k}\, p^k\, (1-p)^{n-k}
	= \frac{n!}{k!(n-k)!}\, p^k\, (1-p)^{n-k}
	\]
	calculates the probability of getting exactly $k$ successes in $n$ trials ($k\le n$) with a probability $p\in [0,1]$ of success in each trial.
	The expected number of counts is
	\[
	\mathbb E[k] = p\, n\, .
	\]
\end{itemize}
\section{Conventional multi-exposure image fusion}
\label{sec:B}
Photon counts on a detector can be assumed to follow a Poisson distribution.
Under the assumption of independent and identically distributed (i.i.d.) counts, the likelihood of acquiring a diffraction pattern $n(\pixel)$ given some true diffraction intensity $\mathcal{I}(\pixel)$ is
\begin{equation}\label{eq:likelihood_noflux}
	\Pr(n(\pixel) \mid \mathcal{I}(\pixel))
	= \poipmf{n(\pixel)}{\mathcal{I}(\pixel)}
	= \frac{[\mathcal{I}(\pixel)]^{n(\pixel)}}{n(\pixel)!} e^{- \mathcal{I}(\pixel)}\, .
\end{equation}
In MEF, we measure diffraction patterns $n_i(\pixel)$ by increasing the optical intensity by some flux factor $c_i$ which is proportional to the acquisition times $t_i$.
Under the assumption that $c_i \approx t_i$, we can rewrite the likelihood of observing $n_i(\pixel)$ counts at pixel $\pixel$ as
\begin{equation}\label{eq:likelihood}
	\Pr(n_i(\pixel) \mid \mathcal{I}, c_i)
	= \poipmf{n_i(\pixel)}{c_i \mathcal{I}(\pixel)}
	= \frac{[c_i \mathcal{I(\pixel)}]^{n_i(\pixel)}}{n_i(\pixel)!} e^{-c_i \mathcal{I}(\pixel)}\, .
\end{equation}
We will occasionally drop the pixel dependence $\pixel$ for convenience.
The log likelihood $\ell_i$ for the fused image $\mathcal I$ resulting from the $i$-th acquition can be written as
\begin{equation}\label{eq:log-likelihood}
	\ell_i(\mathcal{I}) = \displaystyle \sum_{\pixel} \big\{ n_i(\pixel) \log[c_i \mathcal{I}(\pixel)] - c_i \mathcal{I}(\pixel)\big\} \, .
\end{equation}
By incorporating only unsaturated pixels via the pixel-wise weight factor $w_i(\pixel)$, and summing over diffraction patterns $n_i(\pixel)$, the total log-likelihood becomes, 
\begin{equation}
	\ell(\mathcal{I}) = \displaystyle \sum_i \sum_{\pixel} \big\{ w_i(\pixel) n_i(\pixel) \log[c_i \mathcal{I}(\pixel)] - w_i(\pixel) c_i\mathcal{I}(\pixel) \big\} \, .
\end{equation}
The maximum likelihood estimate (MLE) of $\mathcal I$ is obtained by maximizing the log likelihood:
\begin{equation*}
	\mathcal{I}_{\text{MLE}} = \underset{\mathcal{I}}{\operatorname*{arg \,max}}\, \ell(\mathcal{I})\, .
\end{equation*}
The MLE can be computed by setting the gradient of $\ell$ to zero:
\[
\frac{\partial \ell(\mathcal I)}{\partial I(\pixel)} = \sum_i \frac{w_i(\pixel)\, n_i(\pixel)}{\mathcal I(\pixel)} - \sum_i w_i(\pixel) c_i \overset{!}{=} 0\, .
\]
Solving this for $\mathcal I$ the fused MEF result is now given as
\begin{equation*}
	\mathcal{I}_{\text{MLE}}(\pixel) = \frac{\sum_i w_i(\pixel)\,  n_i(\pixel)}{\sum_i w_i(\pixel)\, c_i} \, .
\end{equation*}
After subtracting the dark frame measurements $b_i(\pixel)$ from the diffraction patterns
\begin{equation*}
	n_i^+(\pixel) := \max{\big(0, n_i(\pixel) - b_i(\pixel)\big)} \, ,
\end{equation*}
the conventional MEF result can be rewritten as
\begin{equation}\label{eq:mef_mle2}
	\begin{aligned}
		\mathcal{I}_{\text{MLE}}(\pixel) &= \frac{\sum_i w_i(\pixel)\, n_i^+(\pixel)}{\sum_i w_i(\pixel)\, c_i}, \quad
		w_i(\pixel) = 
		\begin{cases}
			1, & \text{$\pixel$ is not saturated} \\
			0, & \text{$\pixel$ is saturated}
		\end{cases}
	\end{aligned}
\end{equation}
by replacing the raw counts $n_i(\pixel)$ with the background-corrected counts $n_i^+(\pixel)$. 
\section{Bayesian multi-exposure image fusion}
\label{sec:C}
To derive Bayesian multi-exposure image fusion (MEF), we divide it in three parts.
The first part focuses on considering the Poisson likelihood and deriving model parameters for fused data $\mathcal{I}(\pixel)$ and $c_i$.
In the next subsection, we assume that the Poisson data is corrupted with background, therefore modifying our likelihood to accommodate for this.
We derive the model parameters under this situation.
In the last subsection, we assume a {\em right-(censored)} Poisson distribution to model the saturated pixels along with the consideration of the background data.
\subsection{Poisson data}
\label{sec:C1}
The likelihood of observing $n_i(\pixel)$ at pixel $\pixel$ is defined in Eq.~\ref{eq:likelihood}.
The likelihood over the entire dataset can be given as
\begin{equation}\label{eq:full-likelihood}
	\mathcal{L}(\mathcal I, c)
	= \prod_i \prod_{\pixel} \left[c_i \mathcal{I}(\pixel)\right]^{n_i(\pixel)} e^{-c_i \mathcal{I}(\pixel)}\, .
\end{equation}
We now define prior distributions. For $c_i$, we assume a Gamma distribution as prior
\begin{equation}\label{eq:prior-fluxes}
	\Pr(c_i\mid\alpha_i, \beta_i) = \gammapdf{c_i}{\alpha_i}{\beta_i}
	= \frac{\beta_i^{\alpha_i}}{\Gamma(\alpha_i)} \, c_i^{\alpha_i - 1} e^{-\beta_i c_i} 
\end{equation}
with the following properties
\begin{equation}\label{eq:mean-var-prior}
	\mathbb{E}[c_i] = \frac{\alpha_i}{\beta_i}, \quad \text{var}[c_i] = \frac{\alpha_i}{\beta_i^2}\, .
\end{equation}
Similarly, our prior for $\mathcal{I}(\pixel)$ is
\begin{equation}\label{eq:prior-lambda}
	\Pr(\mathcal{I}(\pixel) \mid \alpha_{\mathcal{I}}, \beta_{\mathcal{I}})
	= \gammapdf{\mathcal{I}(\pixel)}{\,\alpha_\mathcal{I}}{\beta_\mathcal{I}}
	= \frac{\beta_\mathcal{I}^{\alpha_\mathcal{I}}}{\Gamma(\alpha_\mathcal{I})} \left[\mathcal{I}(\pixel) \right]^{\alpha_\mathcal{I} - 1} e^{-\beta_\mathcal{I} \mathcal{I}(\pixel)}
\end{equation}
The joint posterior is obtained by invoking Bayes' rule:
\begin{equation}\label{eq:joint-posterior}
	\Pr(\mathcal{I}(\pixel), c_i \mid n_i)
	\propto \Pr(n_i(\pixel) \mid \mathcal{I}(\pixel), c_i) \,
	\Pr(\mathcal{I}(\pixel) \mid \alpha_{\mathcal{I}}, \beta_{\mathcal{I}}) \,
	\Pr(c_i \mid \alpha_i, \beta_i)
\end{equation}
where the dependence on the hyperparameters $\alpha_{\mathcal{I}}, \beta_{\mathcal{I}}, \alpha_i, \beta_i$ is omitted.
Due to the conjugacy property, the conditional posterior distributions for $\mathcal I$ and $c_i$ will also be Gamma distributions.
For $\Pr(\mathcal{I}, \mid c_i, n_i)$, we have
\begin{equation}\label{eq:conditional-posterior-I}
	\Pr(\mathcal{I}(\pixel) \mid c_i, n_i)
	= \gammapdf{\mathcal{I}(\pixel)}{\,\alpha_\mathcal{I} + \sum_i n_i(\pixel)}{\beta_\mathcal{I} + \sum_i c_i}
\end{equation}
Similarly for $\Pr(c_i \mid \mathcal{I}, n_i)$, we have
\begin{equation}\label{eq:conditional-posterior-II}
	\Pr(c_i \mid \mathcal{I}, n_i)
	= \gammapdf{c_i}{\,\alpha_i + \sum_{\pixel} n_i(\pixel) }{\beta_i + \sum_{\pixel} \mathcal{I}(\pixel)}
\end{equation}
Therefore, using the analytical expression for the expectation of the Gamma distribution, the model parameters can be estimated as,
\begin{equation}\label{eq:ICE_updates}
	\begin{aligned}
		\mathcal{I}^{(k+1)}(\pixel)
		&= \mathbb E[\mathcal{I}(\pixel) \mid c_i^{(k)}]
		= \frac{\alpha_\mathcal{I} + \sum_i n_i(\pixel)}{\beta_\mathcal{I} + \sum_i c_i^{(k)}}\, \, ,\\
		c_i^{(k+1)}
		&= \mathbb E[c_i \mid \mathcal{I}^{(k+1)}]
		= \frac{\alpha_i + \sum_{\pixel} n_i(\pixel)}{\beta_i + \sum_{\pixel} \mathcal{I}^{(k+1)}(\pixel)}
	\end{aligned}
\end{equation}
The expectation for a uniform Gamma prior of $\mathcal{I}(\pixel)$ can be given as $\alpha_\mathcal{I}/\beta_\mathcal{I} = 1$.
To achieve this, we simply provide the same value for $\alpha_\mathcal{I}$ and $\beta_\mathcal{I}$.
Here, we choose both of these values to be $10^{-3}$, which also avoids instability in the division.
Since, we expect $c_i \approx t_i$, the prior expectation is then given as $\alpha_i / \beta_i = t_i$.
For the special case of $\alpha_i=1$ which is the exponential distribution, $\beta_i=t_i^{-1}$.
\subsection{Poisson data with background}
\label{sec:C2}
To account for the background or readout noise, we assume the Poisson rate $c_i \mathcal{I}(\pixel)$ is shifted by some background rate $b_i(\pixel)$
\begin{equation}\label{eq:background-model}
	n_i(\pixel) \sim \poipmf{n_i(\pixel)}{c_i\mathcal{I}(\pixel) + b_i(\pixel)}\, .
\end{equation}
The model can be interpreted as observing the sum of the latent unobserved noisefree diffraction patterns $n_{\mathcal{I} i}(\pixel)$ and background counts $n_{bi}(\pixel)$.
\begin{align}\label{eq:background-sum}
	\begin{split}
		&\poipmf{n_i(\pixel)}{c_i\mathcal{I}(\pixel) + b_i(\pixel)} \\
		&= \displaystyle
		\sum_{n_{\mathcal{I} i}(\pixel), n_{bi}(\pixel)=0}^\infty
		\iverson{n_i(\pixel) = n_{\mathcal{I} i}(\pixel) + n_{b i}(\pixel)}\,
		\poipmf{n_{\mathcal{I} i}(\pixel)}{c_i\mathcal{I}(\pixel)}
		\poipmf{n_{b i}(\pixel)}{b_i(\pixel)}\, .
	\end{split}
\end{align}
This follows from the binomial theorem as
\begin{align*}
	\sum_{k, l=0}^{\infty} \iverson{s=k+l}\, \poipmf{k}{x}\, \poipmf{l}{y}
	&=
	\sum_{k, l=0}^{\infty} \iverson{s=k+l}\, \frac{x^k}{k!}\, \frac{y^l}{l!} e^{-(x+y)} \\
	&=
	e^{-(x+y)} \sum_{k=0}^s \frac{x^k}{k!}\, \frac{y^{s-k}}{(s-k)!} \\
	&=
	\frac{e^{-(x+y)}}{s!} \sum_{k=0}^s \binom{s}{k} x^k\, y^{s-k} \\
	&=
	\frac{e^{-(x+y)}}{s!} (x+y)^s = \poipmf{s}{x+y}
\end{align*}
To derive the conditional posterior of the latent counts, we note that the distribution $n_{\mathcal{I} i}(\pixel)$ given $n_i(\pixel)$ is
\begin{equation*}
	\Pr(n_{\mathcal{I} i}(\pixel) \mid n_i(\pixel), \mathcal{I}(\pixel), c_i, b_i(\pixel))
	= \frac{\Pr(n_i(\pixel), n_{\mathcal{I} i}(\pixel) \mid \mathcal{I}(\pixel), c_i, b_i(\pixel))}{\Pr(n_i(\pixel) \mid \mathcal{I}(\pixel), c_i, b_i(\pixel))} 
\end{equation*}
We have for $\Pr(n_i(\pixel), n_{\mathcal{I} i}(\pixel) \mid \mathcal{I}(\pixel), c_i, b_i(\pixel))$
\begin{equation}\label{eq:joint-background-model}
	\begin{aligned}
		\Pr(n_i, n_{\mathcal{I} i} \mid \mathcal{I}, c_i, b_i)
		&= \displaystyle \sum_{n_{bi}=0}^\infty
		\iverson{n_i = n_{\mathcal{I} i} + n_{b i}}\,
		\poipmf{n_{\mathcal{I} i}}{c_i\mathcal{I}}
		\poipmf{n_{b i}}{b_i} \\
		&= \frac{e^{-(c_i\mathcal{I} + b_i)}}{n_i!} \binom{n_i}{n_{\mathcal{I} i}} \left(c_i\mathcal{I}\right)^{n_{\mathcal{I} i}} b_i^{n_i-n_{\mathcal{I} i}} ,
	\end{aligned}
\end{equation}
and $\Pr(n_i(\pixel) \mid \mathcal{I}(\pixel), c_i, b_i(\pixel))$ 
\begin{equation}\label{eq:background-model-derived}
	\Pr(n_i \mid \mathcal{I}, c_i, b_i)
	= \poipmf{n_i}{c_i \mathcal{I} + b_i} 
	= \frac{\left(c_i\mathcal{I} + b_i\right)^{n_i}}{n_i!}e^{-\left(c_i\mathcal{I} + b_i \right)} \, ,
\end{equation}
From Eq.~\ref{eq:joint-background-model} and Eq.~\ref{eq:background-model-derived}, it follows that
\begin{equation}\label{eq:noisefree-data-model}
	\begin{aligned}
		\Pr(n_{\mathcal{I} i} \mid n_i, \mathcal{I}, c_i, b_i)
		&= \frac{e^{-(c_i\mathcal{I} + b_i)}}{n_i!} \binom{n_i}{n_{\mathcal{I} i}}
		\left(c_i\mathcal{I}\right)^{n_{\mathcal{I} i}} b_i^{n_i-n_{\mathcal{I} i}} \biggl/\biggr.
		\frac{\left(c_i\mathcal{I} + b_i\right)^{n_i}}{n_i!}e^{-\left(c_i\mathcal{I} + b_i \right)}  \\
		&= \binom{n_i}{n_{\mathcal{I} i}} \frac{\left(c_i\mathcal{I}\right)^{n_{\mathcal{I} i}}
			b_i^{n_i-n_{\mathcal{I} i}}}{(c_i\mathcal{I}+b_i)^{n_i}}\\
		&= \text{Binomial}\left(n_{\mathcal{I} i}; n_i,
		\frac{c_i\mathcal{I}}{c_i\mathcal{I}+b_i}\right)
	\end{aligned}
\end{equation}
We record the aggregate counts $n_i(\pixel)$ and if we were to measure dark frames  $m_{ij}(\pixel)$, i.e., repeat $J$ measurements per acquisition, we can write it as
\begin{equation*}
	m_{ij}(\pixel) \sim \poipmf{\,\cdot\,}{b_i(\pixel)}, \quad j=1, \ldots, J
\end{equation*}
We can assume the mean of the dark frames to be $\overline{m}_i(\pixel) \approx b_i(\pixel)$ and therefore the background rate $b_i(\pixel)$ to be known for all acquisitions.
We can summarize all the conditional posterior distributions as
\begin{align}\label{eq:conditional-background-posterior}
	\begin{split}
		\mathcal{I}(\pixel)
		&\sim \gammapdf{\mathcal{I}(\pixel)}{\,\alpha_\mathcal{I} + \sum_i n_{\mathcal{I} i} (\pixel)}{\beta_\mathcal{I} + \sum_i c_i} \\
		n_{\mathcal{I} i}(\pixel) &\sim \text{Binomial}\left(n_{\mathcal{I} i}(\pixel); n_i(\pixel), \frac{c_i\mathcal{I}(\pixel) }{ c_i \mathcal{I}(\pixel) + b_i(\pixel)} \right) 
	\end{split}
\end{align}
We can employ the expectation-maximization (EM) algorithm. In the expectation step (E-step), we find the expectation of the latent quantity $n_{\mathcal{I} i}(\pixel)$.
In the maximization step (M-step), we maximize our estimation of the fused diffraction pattern $\mathcal{I}(\pixel)$.
\begin{equation}\label{eq:EM-background}
	\begin{aligned}
		n_{\mathcal{I} i}^{(k+1)}(\pixel)
		&= \mathbb E[n_{\mathcal{I} i}(\pixel) \mid \mathcal{I}^{(k)}(\pixel)]
		=  \frac{c_i\mathcal{I}^{(k)}(\pixel)}{c_i\mathcal{I}^{(k)}(\pixel) + b_i(\pixel)}\,
		n_i(\pixel) \\
		\mathcal{I}^{(k+1)}(\pixel)
		&= \mathbb E[\mathcal{I}(\pixel) \mid n_{\mathcal{I} i}^{(k+1)}(\pixel)]
		= \frac{\alpha_\mathcal{I} + \sum_i n_{\mathcal{I} i}^{(k+1)}(\pixel)}{\beta_\mathcal{I} + \sum_i c_i}
	\end{aligned}
\end{equation}
\subsection{Censored Poisson data with background}
\label{sec:C3}
Now assume that the data are truncated at some {\em censoring} threshold $\nmax$ such that the observed counts $n_i(\pixel)$ are restricted to the range $[0, \nmax]$.
These data can be modeled as unrestricted counts $\nu_i(\pixel) \in \mathbb N$
\begin{equation}\label{eq:censored-model-background}
	\nu_i(\pixel) \sim \poipmf{\,\cdot\,}{c_i \mathcal{I}(\pixel) + b_i(\pixel)}
\end{equation}
that are only partially observed. 
If we knew all $\nu_i(\pixel)$, then the estimation of $\mathcal I(\pixel)$ and $c_i$ could proceed as in the previous section \ref{sec:C2}. 
At saturated pixels (i.e. whenever $\nu_i(\pixel) \ge \nmax$), the observed data are censored such that the model for the actual observations $n_i(\pixel)$ given $\nu_i(\pixel)$ is:
\begin{equation}\label{eq:censored-model-background-condition}
	n_i(\pixel) = \begin{cases}
		\nu_i(\pixel), & \nu_i(\pixel) < \nmax \\
		\nmax, & \nu_i(\pixel) \ge \nmax
	\end{cases}\, .
\end{equation}
Since $\nu_i(\pixel)$ are only partially observed, we estimate them from $n_i(\pixel)$ using their conditional posterior
\begin{equation}\label{eq:censored-posterior}
	\nu_i(\pixel) \sim  \kpoipmf{\,\cdot\,}{c_i \mathcal I(\pixel) + b_i(\pixel)}{\nmax-1}
\end{equation}
when necessary, i.e. when $w_i(\pixel) = 0$ where the binary weight $w_i(\pixel) = \iverson{n_i(\pixel) < \nmax}$ indicates if a pixel is not saturated (when $w_i(\pixel) = 1$, we just let $\nu_i(\pixel) = n_i(\pixel)$).
Here, we introduced the truncated Poisson distribution $\kpoipmf{\nu_i(\pixel)}{c_i \mathcal{I}(\pixel) + b_i(\pixel)}{\nmax-1}$ that corresponds to the unobserved right tail of the Poisson distribution $\poipmf{\nu_i(\pixel)}{c_i \mathcal{I}(\pixel) + b_i(\pixel)}$.
The $k$-truncated Poisson distribution is defined as
\begin{equation}\label{eq:k-truncated-poisson}
	\kpoipmf{n}{\lambda}{k}
	= \frac{\poipmf{n}{\lambda}}{1 - \sum_{j=0}^k \poipmf{j}{\lambda}}
	= \frac{1}{n!}\, \frac{\lambda^n e^{-\lambda}}{1 - e^{-\lambda} \sum_{j=0}^k \lambda^j / j!}
\end{equation}
for $n=k+1, k+2, \ldots $
The expected number of counts under the truncated Poisson distribution is
\begin{equation}\label{eq:mean-truncated}
	\begin{aligned}
		\mathbb E[n \mid \lambda, k]
		&= \sum_{n\ge 0} n\, \kpoipmf{n}{\lambda}{k} 
		= \sum_{n\ge k+1} n \, \frac{\lambda^n e^{-\lambda} / n!}{1 - e^{-\lambda} \sum_{m=0}^k \lambda^m / m!} 
		= \frac{\sum_{n\ge k+1} n \lambda^n e^{-\lambda} / n!}{1 - e^{-\lambda} \sum_{m=0}^k \lambda^m / m!} \\
		&= \frac{\lambda\sum_{n\ge k+1} \lambda^{n-1} e^{-\lambda} / (n-1)!}{1 - e^{-\lambda} \sum_{m=0}^k \lambda^m / m!}
		= \lambda \frac{\sum_{n\ge k} \lambda^{n} e^{-\lambda} / n!}{1 - e^{-\lambda} \sum_{m=0}^k \lambda^m / m!}
		= \lambda\, \frac{1 - e^{-\lambda} \sum_{n=0}^{k-1} \lambda^n / n!}{1 - e^{-\lambda} \sum_{m=0}^k \lambda^m / m!} \, .
	\end{aligned}
\end{equation}
We estimate the unobserved counts $\nu_i(\pixel)$ for $w_i(\pixel) = 0$ by computing their expected value given the current estimate of the rate $c_i\mathcal{I}(\pixel) + b_i(\pixel)$.
Therefore, the expectation of the latent counts $\nu_i(\pixel)$ can be written as
\begin{align}\label{eq:mean-latent-counts}
	\begin{split}
		& \mathbb E\left[\nu_i(\pixel) \mid c_i \mathcal{I}(\pixel) + b_i(\pixel), \nmax - 1 \right] \\
		&= \left(c_i \mathcal{I}(\pixel) + b_i(\pixel) \right) \, \frac{1 - e^{-\left(c_i \mathcal{I}(\pixel) + b_i(\pixel) \right)} \sum_{m=0}^{\nmax - 2} [c_i \mathcal{I}(\pixel) + b_i(\pixel)]^m / m!}{1 - e^{-\left(c_i \mathcal{I}(\pixel) + b_i(\pixel) \right)} \sum_{m=0}^{\nmax - 1} [c_i \mathcal{I}(\pixel) + b_i(\pixel)]^m / m!}
	\end{split}
\end{align}
The expectation maximization (EM) algorithm is now employed to estimate the expectation of our incomplete data and maximize for the fused result $\mathcal{I}(\pixel)$.
The expectation step (E-step) can be summarized as
\begin{equation}
	\nu_i^{(k+1)}(\pixel) = \begin{cases}
		\mathbb{E}[\nu_i(\pixel) \mid c_i \mathcal{I}^{(k)}(\pixel) + b_i(\pixel), \nmax - 1], &
		w_i(\pixel) = 0\\
		n_i(\pixel), & w_i(\pixel) = 1\\
	\end{cases}
\end{equation}
The expectation also includes the estimation of {\em complete} noisefree counts $\nu_{\mathcal{I} i}$ similar to the estimate of $n_{\mathcal{I} i}$ from Eq.~\ref{eq:EM-background}:
\begin{equation}\label{eq:e-step-noisefree-censored-background}
	\nu_{\mathcal{I} i}^{(k+1)}(\pixel) =
	\frac{c_i\mathcal{I}^{(k)}(\pixel)}{c_i\mathcal{I}^{(k)}(\pixel) +
		b_i(\pixel)}\,\nu_i^{(k+1)}(\pixel)
\end{equation}
By assuming Gamma prior $\Pr(\mathcal{I}(\pixel))$ from Eq.~\ref{eq:prior-lambda}, we can maximize the log posterior with respect to $\mathcal{I}(\pixel)$.
The maximization step based on noisefree counts estimates $\mathcal{I}(\pixel)$ as
\begin{equation}\label{eq:m-step-censored-background}
	\begin{aligned}
		&\mathcal{I}^{(k+1)}(\pixel) 
		= \frac{\alpha_\mathcal{I} + \displaystyle \sum_i \nu_{\mathcal{I} i}^{(k+1)}(\pixel)}{\beta_\mathcal{I} + \displaystyle \sum_i c_i} \, . \\
	\end{aligned}
\end{equation}
The full EM algorithm can now be summarized with Eq.~\ref{eq:ICE_updates} and Eq.~\ref{eq:m-step-censored-background} as
\begin{equation}
	\begin{aligned}
		\nu_{\mathcal{I} i}^{(k+1)}(\pixel) &=
		\frac{c_i^{(k)}\mathcal{I}^{(k)}(\pixel)}{c_i^{(k)}\mathcal{I}^{(k)}(\pixel) + b_i(\pixel)}
		\left(w_i(\pixel) n_i(\pixel) + \left(1 - w_i(\pixel) \right)\, \mathbb E[\nu_i(\pixel) | c_i^{(k)} \mathcal{I}^{(k)}(\pixel) + b_i(\pixel), \nmax -1]\right) \\
		c_i^{(k+1)} &= \frac{\alpha_i + \sum_{\pixel} \nu_{\mathcal{I} i}^{(k+1)}(\pixel)}{\beta_i + \sum_{\pixel} \mathcal{I}^{(k)}(\pixel)} \\
		\mathcal{I}^{(k+1)}(\pixel) &= \frac{\alpha_\mathcal{I} + \sum_i \nu_{\mathcal{I} i}^{(k+1)}(\pixel)}{\beta_\mathcal{I} + \sum_i c_i^{(k+1)}}
	\end{aligned}
\end{equation}
\section{Synthetic data}
\label{sec:D}
The section consists of additional details about the SNR of weakly scattering specimens and the structural similarity index (SSIM).

\subsection{Weak phase object approximation}
\label{sec:D1}

A weak phase object can be represented by a superposition of weak phase gratings using Fourier analysis. For a simple weak phase grating,
\begin{equation}
	t(x) = \exp\big(2 \,i \,\rho \cos(2\pi f_0 x)\big)
\end{equation}
where, $\rho\ll1$ is a small constant, $f_0$ is a fixed frequency with $x$ being the spatial coordinate. The Taylor expansion of this becomes,
\begin{equation}
	\begin{aligned}
		t(x) &\approx 1 + 2i\rho \cdot \cos(2\pi f_0 x) + \dots \\
		& = 1 + i\rho \cdot \big(e^{i2\pi f_0 x} + e^{-i2\pi f_0 x} \big) 
	\end{aligned}
\end{equation}
Fourier transform $T(f) = \mathcal{F}\big(t(x)\big)$ can be given as
\begin{equation}\label{eq:ft-object}
	\begin{aligned}
		T(f) = \delta(f) + i\rho \big(\delta(f_0 - f) + \delta(f_0 + f)\big) 
	\end{aligned}
\end{equation}
where, $\delta(f)$ is a delta function. In ptychography, the exit wave is given as
\begin{equation}
	e(x) = t(x) \cdot p(x)
\end{equation}
where $p(x)$ is the probe. Far-field diffraction pattern records Fourier transformed signal which is the convolution between the object and probe in the reciprocal space. This is given as
\begin{equation}\label{eq:far-field}
	E(f) = T(f) \circledast P(f)
\end{equation}
Now plugging Eq.~\ref{eq:ft-object} in Eq.~\ref{eq:far-field}, we can rewrite this as
\begin{equation}\label{eq:full-far-field}
	\begin{aligned}
		E(f) &= \bigg(\delta(f) + i\rho \big(\delta(f_0 - f) + \delta(f_0 + f)\big) \bigg) \circledast P(f) \\
		&= P(f) + i\rho \big(P(f_0 - f) + P(f_0 + f)\big)
	\end{aligned}
\end{equation}
The measured intensity $I(f) = |E(f)|^2$. We can now write simplified $I(f)$ as
\begin{equation}
	I(f) \approx |P(f)|^2 + \rho^2\big|P(f_0 - f) + P(f_0 + f)\big|^2
\end{equation}
Here, $|P(f)|^2$ corresponds to the empty probe beam or ballistic (unscattered) photons as recorded on the detector. Poisson noise is mainly driven by the probe. The second term depends on the object quality which is the phase grating modulation $\rho$. If $C = P(f)$, we can say,
\begin{equation}
	\begin{aligned}
		I_{ballistic} &= C^2 \\
		I_{scattered} &= \rho^2 C^2		
	\end{aligned}
\end{equation}
The noise in the CMOS sensor scales as the square root of the signal $\sqrt{I_{ballistic}}$. Therefore, the signal-to-noise ratio (SNR) can now be given as,
\begin{equation}
	\begin{aligned}
		\text{SNR} &= \frac{I_{scattered}}{\sqrt{I_{ballistic}}} \\
		&= \rho^2 C
	\end{aligned}
\end{equation}
So, in general, SNR scales with the phase modulation term $\rho^2$ of the specimen and the total signal strength $C$. Therefore, a small value of $\rho$, i.e., a weak scattering specimen, leads to a poor SNR. In general, a weak phase object $t(x)$ can be given as 
\begin{equation}
	t(x) = \exp(i\phi) \approx 1 + i\phi
\end{equation}
where $\rho=\max\big(\phi(x)\big)$
\subsection{Structural Similarity index}
\label{sec:D2}
Structural Similarity (SSIM) index~\cite{Wang2004} is defined between windows (spatial patches of the same size) $\textbf{x}$ from a reference image $\textbf{X}$ and $\textbf{y}$ from a test image $\textbf{Y}$ as
\begin{equation}\label{eq:ssim}
	\text{SSIM}(\textbf{x}, \textbf{y}) = \frac{(2\mu_x \mu_y + C_1)(2\sigma_{xy} + C_2)}{(\mu_x^2 + \mu_y^2 + C_1)(\sigma_x^2 + \sigma_y^2 + C_2)}
\end{equation}
This gives us an estimate of local statistics within the window. In this paper, we use a $9 \times 9$ window using a circular symmetric Gaussian weighting function $\textbf{w} = \{ w_i | i = 1, 2, ... , N \}$ with the standard deviation of $1.0$, normalized to unit sum $(\sum_{i=1}^{N}w_i=1)$. The estimates of $\mu_x$ and $\sigma_x^2$ are therefore given as weighted mean and weighted variance respectively.
\begin{equation}
	\begin{aligned}
		\mu_x &= \sum_{i=1}^{N} w_i x_i \\
		\sigma_x^2 &= \sum_{i=1}^{N} w_i (x_i - \mu_x)^2
	\end{aligned}
\end{equation}
This is similar for $\mu_y$ and $\sigma_y^2$. The weighted covariance $\sigma_{xy}$ is given as
\begin{equation}
	\sigma_{xy} = \sum_{i=1}^{N} w_i(x_i - \mu_x)(y_i - \mu_y)
\end{equation}
$C_1 = (K_1L)^2$ and $C_2 = (K_2L)^2$ are constants that avoid instability when either $(\mu_x^2 + \mu_y^2)$ or $(\sigma_x^2 + \sigma_y^2)$ approach zero. $L$ is the dynamic range a pixel value can take which is set to be the difference between the maximum and minimum value from our reference and test image $L = \max(\textbf{X}, \textbf{Y}) - \min(\textbf{X}, \textbf{Y})$. As per the recommendation of the original publication from Wang et al.,~\cite{Wang2004} we set the values of $K_1=0.01$ and $K_2=0.03$. Although these values are somewhat arbitrary, the variations in these values are insensitive to the performance of the SSIM index. If there are $M$ windows that cover our entire image $\textbf{X}$ or $\textbf{Y}$, a moving window average of SSIM is evaluated called the mean SSIM (MSSIM).
\begin{equation}
	\text{MSSIM}(\textbf{X}, \textbf{Y}) = \frac{1}{M}\sum_{j=1}^{M}\text{SSIM}(\textbf{x}_j, \textbf{y}_j)
\end{equation}
For our synthetic data tests, we evaluate MSSIM between ground truth and results from the Bayesian and conventional MEF methods. It is evaluated for just the phase values from the complex-valued ptychographic reconstructions as we simulate a weak phase object. MSSIM metric was utilized with its implementation from the package $\emph{scikit-image}$~\cite{van2014scikit}.
\begin{figure}
	\centering
	\includegraphics[width=0.85\linewidth]{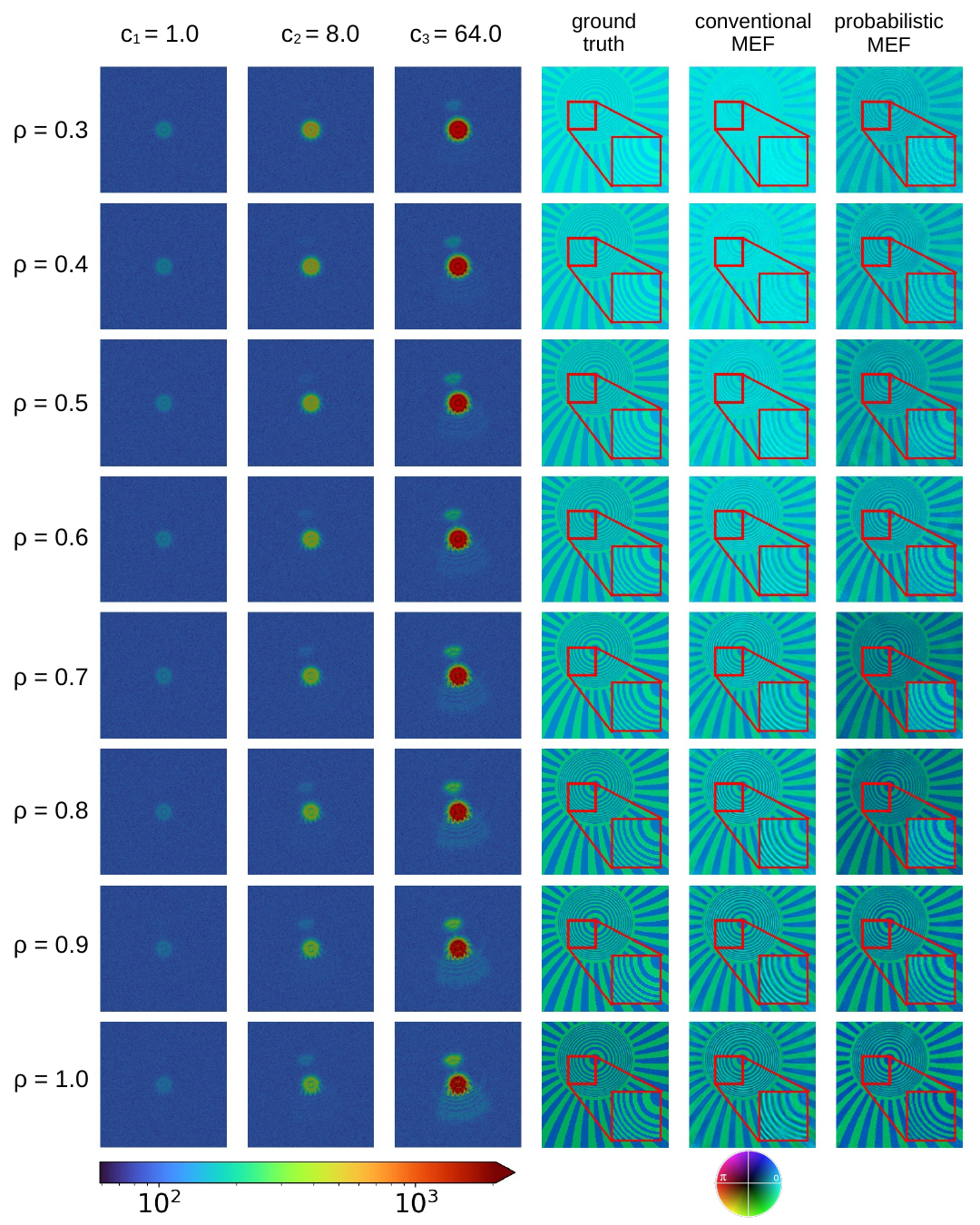}
	\caption{Every row corresponds to synthetic data (log-scale) with ptychographic reconstruction results for scattering parameter $\rho$. The diffraction patterns recorded at varying flux factors $c_i$ are fused with conventional and Bayesian MEF methods and the ptychographic reconstruction results are compared with the ground truth.}
	\label{fig:synthetic-data}
\end{figure}

\section{Experimental design}
\label{sec:E}

\begin{figure}[ht]
	\centering
	\includegraphics[width=0.8\linewidth]{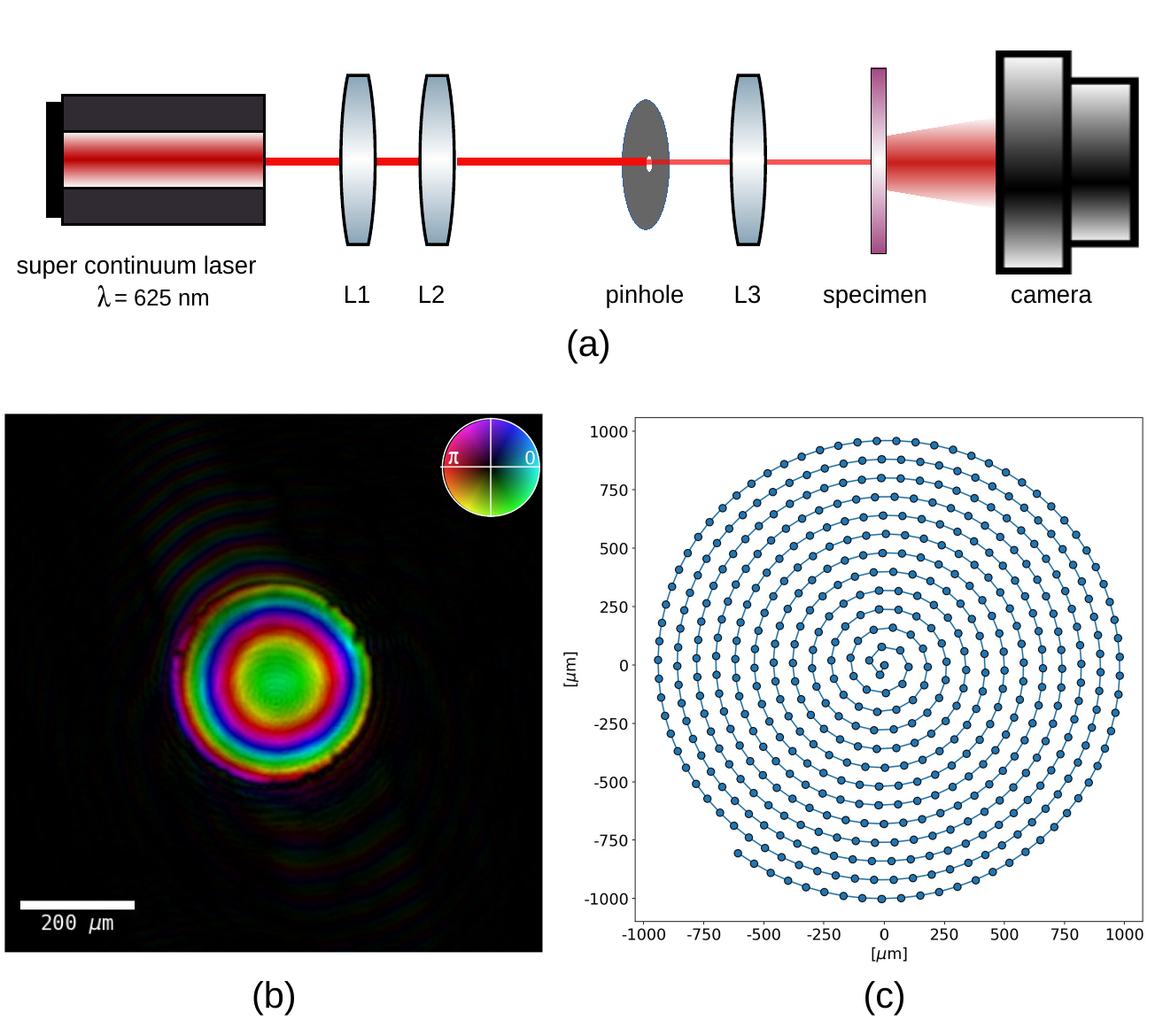}
	\caption{\textbf{(a)} Experimental ptychography setup in transmission geometry. \textbf{(b)} Probe beam of size $400\,\mu m$ characterized with ptychographic reconstruction. \textbf{(c)} Concentric scanning grid used in the experiment.}
	\label{fig:ptychograohy-setup}
\end{figure}

Figure~\ref{fig:ptychograohy-setup} (a) depicts the setup used in this experiment. A coherent light source is generated with the supercontinuum laser (SuperK COMPACT, NKT Photonics) capable of emitting wavelength in a wide range from $450\,nm$ to $2400\,nm$ with a variable repetition rate of $1\,Hz$ to $20\,kHz$. An acousto-optic tunable filter (AOTF) allows wavelength selection with a bandwidth of $2.5\,nm$ to $8.5 \, nm$ from a spectrum of $450\,nm$ to $625\,nm$. For this experiment, we select a wavelength of $625\,nm$ with lens $L1=30\,mm$ and $L2=100\,mm$ used for beam expansion. The beam passes through the fixed pinhole of $400\,\mu m$ which specifies the size of the probe for scanning the specimen. Lens $L3=30\,mm$ images the pinhole onto the specimen. The specimen is mounted on the SmartAct positioner with stages CLS-3232-D-S allowing three degrees of freedom (X, Y, Z direction) driven by a piezo drive. The Z direction is fixed, and the ptychographic scanning is done by moving the specimen in the XY plane. Diffraction patterns are recorded for each scan position with the CMOS camera from Lucid (Phoenix 20 MP Model) with a resolution of $5472 \times 3648 \, px$, pixel size of $2.4\,\mu m \times 2.4\,\mu m$ and a $12$ bit analog-to-digital (ADC) converter. 
In this experiment, we use the unstained histological mouse section as our object. The ptychography experiment is designed by deciding the object-detector distance to be $23.7\,mm$ and a binning of factor $8$ that effectively reduces the resolution to $684 \times 456$. The probe field of view (FOV) can then be calculated to be $\approx 771.4\, \mu m$. Based on the theoretical consideration for the probe to be approximately half of the probe field of view, a pinhole of size $400\,\mu m$ ensures that (see Fig.~\ref{fig:ptychograohy-setup} (b)). We select $500$ scan positions for a concentric spiral scan grid with the radius of $1.009\,mm$ as shown in figure~\ref{fig:ptychograohy-setup} (c) and the probe overlap between adjacent scan positions is $80\,\%$. Therefore, we can reconstruct an area of $\approx 3.2\,mm^2$ from our mouse specimen.


\end{document}